
\def\singlespace{\normalbaselines}
\def\oneandahalfspace{\baselineskip=1.15\normalbaselineskip plus 1pt
\lineskip=2pt\lineskiplimit=1pt}

\def\itemitemitem{\par\indent\indent \hangindent3\parindent \textindent}

\def\np{\vfill\eject}
\def\nl{\hfil\break}

\def\nofirstpagenoten{\nopagenumbers\footline={\ifnum\pageno>1\tenrm
\hss\folio\hss\fi}}
\def\nofirstpagenotwelve{\nopagenumbers\footline={\ifnum\pageno>1\twelverm
\hss\folio\hss\fi}}
\def\leaderfill{\leaders\hbox to 1em{\hss.\hss}\hfill}
\def\ft#1#2{{\textstyle{{#1}\over{#2}}}}
\def\frac#1/#2{\leavevmode\kern.1em
\raise.5ex\hbox{\the\scriptfont0 #1}\kern-.1em/\kern-.15em
\lower.25ex\hbox{\the\scriptfont0 #2}}
\def\sfrac#1/#2{\leavevmode\kern.1em
\raise.5ex\hbox{\the\scriptscriptfont0 #1}\kern-.1em/\kern-.15em
\lower.25ex\hbox{\the\scriptscriptfont0 #2}}

\font\rmb=cmr17                   

\parindent=20pt
\def\narrow{\advance\leftskip by 40pt \advance\rightskip by 40pt}

\def\AB{\bigskip
        \centerline{\bf ABSTRACT}\medskip\narrow}
\def\nonarrower{\advance\leftskip by -40pt\advance\rightskip by -40pt}
\def\AE{\bigskip\nonarrower}

\def\boxit#1{\vbox{\hrule\hbox{\vrule\kern3pt
        \vbox{\kern3pt#1\kern3pt}\kern3pt\vrule}\hrule}}

\def\gtorder{\mathrel{\raise.3ex\hbox{$>$}\mkern-14mu
             \lower0.6ex\hbox{$\sim$}}}
\def\ltorder{\mathrel{\raise.3ex\hbox{$<$}|mkern-14mu
             \lower0.6ex\hbox{\sim$}}}
\def\dalemb#1#2{{\vbox{\hrule height .#2pt
        \hbox{\vrule width.#2pt height#1pt \kern#1pt
                \vrule width.#2pt}
        \hrule height.#2pt}}}

\font\fourteentt=cmtt10 scaled \magstep2
\font\fourteenbf=cmbx12 scaled \magstep1
\font\fourteenrm=cmr12 scaled \magstep1
\font\fourteeni=cmmi12 scaled \magstep1
\font\fourteenss=cmss12 scaled \magstep1
\font\fourteensy=cmsy10 scaled \magstep2
\font\fourteensl=cmsl12 scaled \magstep1
\font\fourteenex=cmex10 scaled \magstep2
\font\fourteenit=cmti12 scaled \magstep1
\font\twelvett=cmtt10 scaled \magstep1 \font\twelvebf=cmbx12
\font\twelverm=cmr12 \font\twelvei=cmmi12
\font\twelvess=cmss12 \font\twelvesy=cmsy10 scaled \magstep1
\font\twelvesl=cmsl12 \font\twelveex=cmex10 scaled \magstep1
\font\twelveit=cmti12
\font\tenss=cmss10
 
 \font\ninebf=cmbx7 scaled \magstep1
\font\ninerm=cmr7 scaled \magstep1 \font\ninei=cmmi7 scaled \magstep1
\font\ninesy=cmsy7 scaled \magstep1 
\font\eightrm=cmr7 scaled 1140 
 
\font\sevenbf=cmbx7 \font\sevenrm=cmr7 \font\seveni=cmmi7
\font\sevensy=cmsy7 

\catcode`@=11
\newskip\ttglue
\newfam\ssfam

\def\fourteenpoint{\def\rm{\fam0\fourteenrm}
\textfont0=\fourteenrm \scriptfont0=\tenrm \scriptscriptfont0=\sevenrm
\textfont1=\fourteeni \scriptfont1=\teni \scriptscriptfont1=\seveni
\textfont2=\fourteensy \scriptfont2=\tensy \scriptscriptfont2=\sevensy
\textfont3=\fourteenex \scriptfont3=\fourteenex \scriptscriptfont3=\fourteenex
\def\it{\fam\itfam\fourteenit} \textfont\itfam=\fourteenit
\def\sl{\fam\slfam\fourteensl} \textfont\slfam=\fourteensl
\def\bf{\fam\bffam\fourteenbf} \textfont\bffam=\fourteenbf
\scriptfont\bffam=\tenbf \scriptscriptfont\bffam=\sevenbf
\def\tt{\fam\ttfam\fourteentt} \textfont\ttfam=\fourteentt
\def\ss{\fam\ssfam\fourteenss} \textfont\ssfam=\fourteenss
\tt \ttglue=.5em plus .25em minus .15em
\normalbaselineskip=16pt
\abovedisplayskip=16pt plus 4pt minus 12pt
\belowdisplayskip=16pt plus 4pt minus 12pt
\abovedisplayshortskip=0pt plus 4pt
\belowdisplayshortskip=9pt plus 4pt minus 6pt
\parskip=5pt plus 1.5pt
\setbox\strutbox=\hbox{\vrule height12pt depth5pt width0pt}
\let\sc=\tenrm
\let\big=\fourteenbig \normalbaselines\rm}
\def\fourteenbig#1{{\hbox{$\left#1\vbox to12pt{}\right.\n@space$}}}

\def\twelvepoint{\def\rm{\fam0\twelverm}
\textfont0=\twelverm \scriptfont0=\ninerm \scriptscriptfont0=\sevenrm
\textfont1=\twelvei \scriptfont1=\ninei \scriptscriptfont1=\seveni
\textfont2=\twelvesy \scriptfont2=\ninesy \scriptscriptfont2=\sevensy
\textfont3=\twelveex \scriptfont3=\twelveex \scriptscriptfont3=\twelveex
\def\it{\fam\itfam\twelveit} \textfont\itfam=\twelveit
\def\sl{\fam\slfam\twelvesl} \textfont\slfam=\twelvesl
\def\bf{\fam\bffam\twelvebf} \textfont\bffam=\twelvebf
\scriptfont\bffam=\ninebf \scriptscriptfont\bffam=\sevenbf
\def\tt{\fam\ttfam\twelvett} \textfont\ttfam=\twelvett
\def\ss{\fam\ssfam\twelvess} \textfont\ssfam=\twelvess
\tt \ttglue=.5em plus .25em minus .15em
\normalbaselineskip=14pt
\abovedisplayskip=14pt plus 3pt minus 10pt
\belowdisplayskip=14pt plus 3pt minus 10pt
\abovedisplayshortskip=0pt plus 3pt
\belowdisplayshortskip=8pt plus 3pt minus 5pt
\parskip=3pt plus 1.5pt
\setbox\strutbox=\hbox{\vrule height10pt depth4pt width0pt}
\let\sc=\ninerm
\let\big=\twelvebig \normalbaselines\rm}
\def\twelvebig#1{{\hbox{$\left#1\vbox to10pt{}\right.\n@space$}}}

\def\tenpoint{\def\rm{\fam0\tenrm}
\textfont0=\tenrm \scriptfont0=\sevenrm \scriptscriptfont0=\fiverm
\textfont1=\teni \scriptfont1=\seveni \scriptscriptfont1=\fivei
\textfont2=\tensy \scriptfont2=\sevensy \scriptscriptfont2=\fivesy
\textfont3=\tenex \scriptfont3=\tenex \scriptscriptfont3=\tenex
\def\it{\fam\itfam\tenit} \textfont\itfam=\tenit
\def\sl{\fam\slfam\tensl} \textfont\slfam=\tensl
\def\bf{\fam\bffam\tenbf} \textfont\bffam=\tenbf
\scriptfont\bffam=\sevenbf \scriptscriptfont\bffam=\fivebf
\def\tt{\fam\ttfam\tentt} \textfont\ttfam=\tentt
\def\ss{\fam\ssfam\tenss} \textfont\ssfam=\tenss
\tt \ttglue=.5em plus .25em minus .15em
\normalbaselineskip=12pt
\abovedisplayskip=12pt plus 3pt minus 9pt
\belowdisplayskip=12pt plus 3pt minus 9pt
\abovedisplayshortskip=0pt plus 3pt
\belowdisplayshortskip=7pt plus 3pt minus 4pt
\parskip=0.0pt plus 1.0pt
\setbox\strutbox=\hbox{\vrule height8.5pt depth3.5pt width0pt}
\let\sc=\eightrm
\let\big=\tenbig \normalbaselines\rm}
\def\tenbig#1{{\hbox{$\left#1\vbox to8.5pt{}\right.\n@space$}}}
\let\rawfootnote=\footnote \def\footnote#1#2{{\rm\parskip=0pt\rawfootnote{#1}
{#2\hfill\vrule height 0pt depth 6pt width 0pt}}}

\def\tenfoot{\tenpoint\hskip-\parindent\hskip-.1cm}

\def\semidirprod{\rlap{{\ss C}}\raise1.2pt\hbox{$\mkern.75mu\times$}}
\def\R{\rlap I\mkern3mu{{\rm R}}}
\def\H{\rlap I\mkern3mu{{\rm H}}}
\def\C{\mkern1mu\raise2.2pt\hbox{$\scriptscriptstyle|$}\mkern-7mu{{\rm C}}}
\def\O{\mkern1mu\raise2.2pt\hbox{$\scriptscriptstyle|$}\mkern-7mu{{\rm O}}}
\def\tr{{\rm tr}}
\def\im{{\rm i}}
\def\oneone{\rlap 1\mkern4mu{\rm l}}
\def\for{\lower6pt\hbox{$\Big|$}}
\def\sm#1{{\scriptstyle{#1}}}
\def\ssm#1{{\scriptscriptstyle{#1}}}
\twelvepoint
\vsize=22cm
\oneandahalfspace
\nofirstpagenotwelve
\rightline{IMPERIAL/TP/90-91/16}
\vskip 2cm
\centerline{\rmb The Superparticle and the Lorentz Group}
\vskip 1.5truecm
\centerline{A.S.\ Galperin$^1$\footnote{$^\dagger$}{\tenfoot On leave from
the Laboratory of Theoretical Physics, Joint Institute for Nuclear Research,
Dubna, Head Post Office P.O. Box 79, 101000 Moscow USSR.}, P.S.\
Howe$^2$ and K.S.\ Stelle$^1$}
\vskip 1.5truecm \itemitemitem{$^1$}{\it The
Blackett Laboratory, Imperial College, London SW7 2BZ\/}
\itemitemitem{$^2$}{\it Department of Mathematics, King's College,
London WC2R 2LS\/}
\vskip 2truecm
\AB
\singlespace
We present a unified group-theoretical framework for superparticle theories.
This explains the origin of the ``twistor-like'' variables that have been
used in trading the superparticle's $\kappa$-symmetry for worldline
supersymmetry. We show that these twistor-like variables
naturally parametrise the coset space ${\cal G}/{\cal H}$, where $\cal G$ is
the
Lorentz group $SO^\uparrow(1,d-1)$ and $\cal H$ is its {\it maximal
subgroup\/}.  This space is a  compact manifold, the sphere $S^{d-2}$. Our
group-theoretical construction gives the proper covariantisation of a fixed
light-cone frame and clarifies the relation between target-space and
worldline supersymmetries.
\AE\oneandahalfspace
\np
\noindent{\bf 1. Introduction}
\medskip
     The Lorentz covariant formulation of the superstring theory and of its
infinite tension limit - the massless relativistic superparticle - is a problem
that has turned out to be more subtle than one might originally have expected.
In special dimensions of the target space-time ($d=3,4,6$ and
$10$), these theories describe equal number of  bosonic and fermionic physical
degrees of freedom.
 The main reason for this subtlety derives from the requirements of
Lorentz invariance, which gives a much larger dimension to the spinor variable
of the superstring/superparticle than to the bosonic coordinate.
This is also a problem in supersymmetric field theories.  The known
ways of resolving this problem involve either auxiliary fields or the presence
of a gauge symmetry to equalise the bosonic and fermionic physical degrees of
freedom (or both). For the superstring/superparticle, this situation
requires the existence of a non-trivial fermionic symmetry [1], which is
generally called $\kappa$-symmetry. Generically, the
$\kappa$-transformation for the spinor variable reads
$$
\delta\theta=\rlap/p\kappa \eqno(1.1)
$$

 Unfortunately, this $\kappa$-symmetry has not yet found a
completely satisfactory formulation in some of the most interesting cases.  The
principal difficulty can be easily understood: the parameter of the
$\kappa$-transformations is a spinor of the same dimension as the spinor
variable, but only a half of it contributes to the l.h.s. of (1.1) owing to the
degeneracy of the momentum $\rlap/p$ (on shell $p^2=\rlap/p\rlap/p=0$).
Moreover,   the $\kappa$-transformations only form a closed algebra subject to
the theory's equations of motion.  One well-known way out of these difficulties
is to fix  a  light-cone frame, in which  spinors become reducible
representations of the surviving ``little'' group and  the
$\kappa$-parameter can be completely fixed. After this step, there are equal
surviving
numbers of bosonic and fermionic degrees of freedom.  However, choosing
the light-cone gauge destroys  manifest Lorentz covariance.
On the other hand,  insisting on a Lorentz
covariant quantization leads to the notorious ``ghosts for ghosts'' difficulty:
in order to fix the irrelevant half of  $\kappa$ one should introduce a spinor
ghost variable, but this has twice as many components as one needs to
 compensate,
so one half of it is in its own turn irrelevant and so on {\it ad infinitum}
(see, for example [2] and refs therein).

     Progress towards a resolution of these problems was made by Sorokin,
Tkach and Volkov [3], who rewrote the superparticle action by introducing a
commuting spinor variable and requiring a local worldline supersymmetry of
the usual form.  Upon use of the equations of motion, this local worldline
supersymmetry can then be re-expressed as a $\kappa$-symmetry
transformation.\footnote{$^{*}$}{\tenfoot It is worthwhile to mention that
the STV theory is different from the so-called spinning superparticle theory
[4]
: {\it i}) they describe different physical degrees of freedom, {\it ii}) the
latter possesses  both  worldline supersymmetry and $\kappa$-symmetry,
while in the former $\kappa$-symmetry is traded for the worldline
supersymmetry.}
 The commuting spinor $\psi$ provides an algebraic solution to the particle's
equation of motion  $p^2=0$ through the relation
$p^m\sim\bar\psi\gamma^m\psi$.  In the cases of the $d=2+1$, $d=3+1$ [3,5,6]
and $d=5+1$ [5, 6] superparticles, this approach yields a manifestly
superPoincar\'e invariant formulation with all symmetries linearly realised (in
fact, these actions possess a rigid superconformal invariance in the target
superspace) and with a closed local worldline algebra.

 In $d=9+1$, however, the problem has proved more subtle still. In ref.\ [7],
an
action for the $d=9+1$ superparticle was proposed  that incorporates some
of the ideas of ref.\ [3].   As in [3], the $p^2=0$ constraint is solved
there by the introduction of a Majorana-Weyl commuting spinor, and after that
the dynamical variables are  redefined  using this spinor.  As in the $d=2+1,
3+1\, \&\, 5+1$ cases, this reformulation introduces a new ``pregauge''
symmetry
that transforms the commuting spinor variable.  In the $d=2+1, 3+1\, \&\, 5+1$
cases, these pregauge transformations  are  linear $Z_2,  U(1), SU(2)$
transformations correspondingly, and are accompanied by a scale
transformation; they all may be
incorporated naturally into the local worldline supergroup.  In the
$d=9+1$ case, however, the pregauge transformation is non-linear:
its parameter is a commuting
Majorana-Weyl spinor, but only  seven of its components are relevant.
This parameter is determined up to a further gauge transformation with a vector
parameter, and the latter is determined up to a scalar, so the
transformation has $16-10+1=7$ independent parameters. Another feature of the
formulation
 of ref.\ [7] is that the off-shell numbers of bosonic and fermionic degrees of
freedom are not equal.

     In this paper, we shall adopt a uniform group-theoretical basis for
the study of all the superparticle cases. We shall identify the commuting
spinor variable $\psi$   as a parameter of a coset space  ${\cal G}/{\cal
H}$, where $\cal G$ is the proper orthochronous Lorentz group
$SO^\uparrow(1,d-1)$ and $\cal H$ is its {\it maximal subgroup}. This
space has dimension $d-2$, it is {\it compact} and corresponds to the
sphere $S^{d-2}$ (which is known as the celestial sphere [8] in the $d=4$
case).  The extension of the ordinary superspace coordinates $x,\theta$
by this sphere is very reminiscent of the harmonic superspace formalism
[9] in which commuting isospinor coordinates taking values in the sphere
$S^2$ are introduced in order to organise infinite sets of auxiliary or
gauge fields via harmonic expansions. In particular, the sphere $S^{d-2}$
may be viewed as the manifold of all possible choices of light-cone
frames, allowing one to have the benefits of a light-cone gauge without
losing Lorentz covariance, much in the same way as in the harmonic
superspace formalism, where  the sphere $S^2$ may be viewed as the
 manifold of all  possible choices of complex structures in $N=2, d=4$
supersymmetry.

The above group-theoretical basis  allows for a consistent understanding
of the superparticle theories for $d=3,4\&6$  constructed in [3,5,6]. It
explains some peculiarities of ref.\ [7] that arise as a result of a  gauge
fixing. It also establishes  relations to the various
approaches to an interpretation of the  $d=10$ supersymmetric Yang-Mills
and supergravity constraints as integrability conditions [10]--[13].

Another motivation of the present paper is to improve on previous
approaches  to   light-cone harmonic superspace, which were based on the
coset space  $SO(1,d-1)/[SO(d-2)\times SO(1,1)]$ [14,15]. The latter has
dimension $2(d-2)$ and it is {\it noncompact}. There is an essential
difference between the compact and noncompact coset spaces of the Lorentz
group:  square-integrable functions on  noncompact coset spaces  contain
in their decomposition only infinite-dimensional representations of the
Lorentz group, while the analogous functions on compact spaces may
contain both finite and infinite-dimensional representations (see, for
example [16]). Since  ordinary physics is based on  finite-dimensional
representations, compact coset spaces are preferred as candidates for
light-cone harmonic superspace.

The present paper is organized as follows. First, we examine a new action
for the ordinary massless particle, which contains a null-vector as a
dynamical variable, and establish its group-theoretical meaning as a
parameter of the smallest possible coset space of the Lorentz group
$SO^\uparrow(1,d-1)$. This coset space can be constructed either in the
vector or in the spinor representation of  $SO^\uparrow(1,d-1)$. The
latter is the  relevant one for the case of the superparticle. In
sections 3,4 and 5 we systematically consider superparticles in $d=3,4,$
and $6$ from this point of view. Section 6 is devoted to the $d=10$ case.
In section 7 we summarize our results  and formulate some open problems;
we also give an appendix devoted to the worldline superconformal groups.
\bigskip
\noindent{\bf 2. The bosonic particle and $S^{d-2}$}
\medskip
     We begin with a discussion of the ordinary bosonic
relativistic particle in $d$ dimensions.  Following [3], we write the
action for the particle as $$
I=\int d\tau \; p_m(\tau)[\dot x^m(\tau)-v^m(\tau)],\eqno(2.1)
$$
where the vector $v^m$ is  constrained to be lightlike,
$$
v^mv_m=0,\eqno(2.2)
$$
and must also satisfy the further covariant condition (with respect to the
orthochronous subgroup $SO^\uparrow(1,d-1)$ of the Lorentz group)
$$
v^0>0.\eqno(2.3)
$$
Note that this inequality is necessitated by the following  physical
requirement: at any point of the worldline, $\tau$ can be interpreted as
a time parameter; that is $\dot x^m(\tau)>0$. Hence
the vector $v^m$ must be strictly {\it nonvanishing}.

 The action (2.1) is invariant under arbitrary worldline
reparametrisations, $\tau\to\tau'(\tau)$  which preserve the worldline
orientation, {\it i.e.}, they are monotonic,  $d\tau'/d\tau>0$.  Note from
(2.1) that $v^m$ must transform as a
worldline density under reparametrisations
$$
v'^m(\tau')={d\tau\over d\tau'}v^m(\tau).\eqno(2.4)
$$
  In order to relate (2.1) to the
more familiar form of the massless particle action, one first incorporates the
constraint (2.2) into the action with a Lagrange multiplier $e^{-1}$,
$$
I=\int d\tau \left(p_m(\tau)[\dot x^m(\tau)-v^m(\tau)] +
\ft12e^{-1}v^mv_m\right),\eqno(2.5)
$$
and then varies with respect to $v^m$ to find $v_m=ep_m$, which may be
substituted back into (2.5) to obtain the familiar massless particle action
$$
I=\int d\tau [p_m\dot x^m-\ft12ep^2].\eqno(2.6)
$$

An interesting feature of the action (2.1) (as compared to (2.6)) is that it is
 covariant under general coordinate transformations without using the einbein
field $e(\tau)$ and thus it is reminiscent of Chern-Simons type actions.

     The vector $v^m$ subject to the constraints (2.2, 2.3)
and
considered modulo the $\tau$ repara\-metrisations (2.4) takes its values in a
sphere $S^{d-2}$. This can be seen  by breaking up $v^m$ into ($v^0$, $v^{\hat
m}$), $\hat m=1,2,\ldots,d-1$, and then using the $\tau$ reparametrisations to
pick the gauge $v^0=1$ (remembering that $v^0$ transforms as a density), so
that
$v^{\hat m}v^{\hat m}=1$.  Another way to see this is to pass over to
gauge-invariant coordinates $\xi^{\hat m}=v^{\hat m}/v^0$, for which $\xi^{\hat
m}\xi^{\hat m}=1$.

     A group-theoretic interpretation of the space in which the
Lorentz vector $v^m$ takes its values is in terms of the coset space
${\cal G}/{\cal H}$,
$$
S^{d-2}={SO^\uparrow(1,d-1)\over[SO(d-2)\times SO^\uparrow(1,1)]\semidirprod
\{K_{\tilde m}\}},\eqno(2.7)
$$
where ${\cal G}=SO^\uparrow(1,d-1)$ is the proper orthochronous Lorentz
group in $d$ dimensions, while the divisor subgroup  ${\cal H}=[SO(d-2)\times
SO^\uparrow(1,1)]\semidirprod \{K_{\tilde m}\}$ is its {\it maximal} proper
subgroup. Here $\tilde m=1,2,\ldots d-2$,
 $\{K_{\tilde m}\}$ denotes
the abelian group generated by the ``conformal boosts'' $K_{\tilde m}$ and
$\semidirprod$ denotes a semidirect product. To clearly understand the
structure of the coset (2.7) it is  very useful  to view the $d$-dimensional
Lorentz group $SO^\uparrow(1,d-1)$ as the  {\it conformal} group of the
Euclidean  $(d-2)$-dimensional space. Then the Lorentz generators $L_{mn}=-
L_{nm}$, which satisfy  the algebra\footnote{$^*$}{\tenfoot We use the
``mostly plus'' convention for the Minkowski tensor:
$\eta_{mn}=diag(-1,1,\ldots 1)$.}
$$
[L_{mn},L_{kl}]=\eta_{mk}L_{nl}-\eta_{ml}L_{nk} - \; (m\leftrightarrow n),
\eqno(2.8)
$$
can be separated into the generators of
$SO(d-2)$ rotations
 $L_{{\tilde m}{\tilde n}}$,  $SO^\uparrow(1,1)$ scale transformations
$D=L_{0,d-1}$, $(d-2)$-dimensional translations
$P_{\tilde m}= L_{0,{\tilde m}}-L_{{\tilde m},d-1}$ and conformal boosts
$K_{\tilde m}= L_{0,{\tilde m}}+L_{{\tilde m},d-1}$. The subgroup $\cal H$ is
then generated  by
 $\{L_{{\tilde m}{\tilde n}},\;\; D,\;\; K_{\tilde m}\}$.

The coset (2.7)
has a number of interesting features: {\it i}) its dimension is $d-2$ and
it is the smallest possible coset of the Lorentz group $\cal G$ . This
corresponds to the fact that the subgroup $\cal H$ is the maximal proper
subgroup of $\cal G$; {\it ii}) both $\cal G$ and $\cal H$ are noncompact
and at the same time the coset space is compact; {\it iii}) the coset
space generators $P_{\tilde m}$ form a maximal {\it lightlike} abelian
subgroup of $\cal G$. This  means that they mutually commute and the
Cartan-Killing norm of these generators, as well as of any of their linear
combinations, vanishes.

In order to see why (2.7) is the space that $v^m$ takes its values in, one can
look for the maximal subgroup of the  Lorentz group  $SO^\uparrow(1,d-1)$
that preserves  a given light-like vector $v^m$ with $v^0>0$,
modulo dilatations of $v^m$. Passing to a light-cone frame, where this vector
has only the first and the last components, $v^m=(v,0,\ldots,0,-v)$, it is easy
to see that the sought-for subgroup coincides with $\cal H$.  For this
reason, the variable $v^m$ (2.2-2.4) parametrizes the space of all
possible light-cone frames, so that its introduction
amounts to covariantising  the light-cone.

It is instructive to give a global description of the coset space (2.7),
starting from a matrix $v^m{}_a$ that is an element of
$SO^\uparrow(1,d-1)$, so that it satisfies
$$\eqalignno{
\eta_{mn}v^m{}_av^n{}_b&=\eta_{ab}&(2.9a)\cr
\det(v^m{}_a)&=1&(2.9b)\cr
v^0{}_0&\ge1.&(2.9c)\cr}
$$
As usual in nonlinear realisations, we define the group action of $\cal G$=
$SO^\uparrow(1,d-1)$ by left multiplication,
$$
v'^m{}_a=\Lambda^m{}_nv^n{}_a,\eqno(2.10)
$$
where $\Lambda^m{}_n\in \cal G$.  On the other hand, in
constructing (2.7),  $v^m{}_a$ must be considered to be defined only up to
multiplication on the right by arbitrary elements of the divisor group $\cal
H$=
$[SO(d-2)\times SO^\uparrow(1,1)]\semidirprod \{K_{\tilde m}\}$
$$
v'^m{}_a=v^m{}_b\Omega^b{}_a \eqno(2.11)
$$
where $\Omega^b{}_a \in \cal H$.

 Now, although the left index $m$ on $v^m{}_a$ corresponds to an irreducible
(vector) representation of $SO^\uparrow(1,d-1)$, the right index $a$
corresponds
to a representation of the divisor group that is reducible, but not completely
reducible, {\it i.e.}\ it can be transformed into block triangular form,
but not into block diagonal form.  It is easiest to see this using a
light-cone basis for the indices $a,b,\ldots$ of $\cal H$  (note that
this step does not break the Lorentz group acting on the indices
$m,n,\ldots$) : $$
v^m{}_\pm=\ft1{\sqrt 2}(v^m{}_0\pm v^m{}_{d-1}).\eqno(2.12)
$$
In this basis, the $[SO(d-2)\times SO^\uparrow(1,1)]\semidirprod
\{K_{\tilde m}\}$ ``gauge transformations'' have the upper-triangular form

$$
(v^m{}_-,v^m{}_{\tilde a},v^m{}_+)'=
\matrix{(v^m{}_-,v^m{}_{\tilde b},v^m{}_+)\cr
\phantom{(v^m{}_-,v^m{}_{\tilde b},v^m{}_+)}\cr
\phantom{(v^m{}_-,v^m{}_{\tilde b},v^m{}_+)}}
\left(\matrix{(\Omega)^{-1}&\Omega^-{}_{\tilde a}
 &\ft12\Omega(\Omega^-{}_{\tilde c})^2\cr
0&\Omega^{\tilde b}{}_{\tilde a}&\Omega\Omega^-{}_{\tilde c}\Omega^{\tilde
b}{}_{\tilde c}\cr
0&0&\Omega\cr}\right) \eqno(2.13)
$$
where $\Omega$ is a positive $SO^\uparrow(1,1)$ scale transformation parameter,
$\Omega^{\tilde b}{}_{\tilde a}\in SO(d-2)$ and $\Omega^-{}_{\tilde b}$
corresponds to the $\{K_{\tilde m}\}$ transformations.

Note that $v^m{}_-$ transforms homogeneously in (2.13)
$$
v'^m{}_-=(\Omega)^{-1} v^m{}_- \eqno(2.14)
$$
while $v^m{}_{\tilde a}$ and $v^m{}_+$
transform into themselves as well as into $v^m{}_-$ (and hence the incomplete
reducibility of the representation).

     The matrix components  $v^m{}_-,v^m{}_{\tilde a}$, and $v^m{}_+$
satisfy the following constraints, which are just (2.9) written in the
light-cone basis (2.11):
$$\eqalignno{
\eta_{mn}v^m{}_-v^n{}_-&=0&(2.15a)\cr
\eta_{mn}v^m{}_+v^n{}_+&=0&(2.15b)\cr
\eta_{mn}v^m{}_+v^n{}_-&=-1&(2.15c)\cr
\eta_{mn}v^m{}_\pm v^n{}_{\tilde a}&=0&(2.15d)\cr
\eta_{mn}v^m{}_{\tilde a}v^n{}_{\tilde b}&=\delta_{\tilde a
\tilde b}&(2.15e)\cr
\det(v^m{}_-,v^m{}_{\tilde a},v^m{}_+)&=1&(2.15f)\cr
v^0{}_-&>0.&(2.15g)\cr}
$$
The condition (2.15g) follows from $\eta^{ab}v^m{}_av^n{}_b=\eta^{mn}$
(conjugate to (2.9a)) and (2.9c).

     In order to see that the manifold (2.7) is $S^{d-2}$, we shall now
consider the first column $v^m{}_-$  and note from (2.14) that owing to the
local $SO^\uparrow(1,1)$ symmetry, this vector is defined only up to a local
scale factor.  At this point, we can establish the link with the
particle's variable $v^m$, which is to be identified with $v^m{}_-$, while
the local $SO^\uparrow(1,1)$ symmetry acting on it is to be identified
with the $\tau$ reparametrisation symmetry (2.4). Being a column of an
orthogonal matrix this vector is naturally nonvanishing. From
(2.15a), we see that $v^m=v^m{}_-$ is a null vector, as required by the
variational principle (2.1),(2.2). Finally, the time component of this vector
is
positive (2.15g) in accordance with (2.3).

 Given $v^m{}_-$, the remaining components $v^m{}_{\tilde a}$ and $v^m{}_+$ are
all eliminable, either by the remaining constraints (2.15b--f) or by the
$SO(d-2)\semidirprod\{K_{\tilde m}\}$ gauge transformations.  The best way
to see how this may be done is to use the   $SO^\uparrow(1,d-1)$
covariance of (2.15) and transform the vector $v^m{}_-$ to a special form
such as $(1,0,\ldots,0,-1)$, in which (2.15a--f) take an especially simple
form that allows one to see that $v^0{}_+$, $v^{d-1}{}_+$, $v^0{}_{\tilde a}$
and $v^{d-1}{}_{\tilde a}$ can all be solved for in terms of $v^{\tilde m}{}_+$
and $v^{\tilde m}{}_{\tilde a}\in SO(d-2)$.  These may then be gauged away by
the $\{K_{\tilde m}\}$ and the $SO(d-2)$ transformations, with
parameters $\Omega^-{}_{\tilde a}$ and $\Omega^{\tilde b}{}_{\tilde a}$
(2.13) respectively.

Thus we have shown that the constraints (2.15) for the space (2.7), defined
in terms of the $v^m{}_a$  modulo the divisor group $[SO(d-2)\times
SO^\uparrow(1,1)]\semidirprod \{K_{\tilde m}\}$, give the same manifold as that
defined by a vector $v^m$ that is required to satisfy (2.2, 2.3) and defined
modulo reparametrisations (2.4).  As we have showed earlier, this space is the
sphere $S^{d-2}$.

     The Lorentz-covariant parametrisation of the sphere $S^{d-2}$ by the
matrix $v^m{}_a$ differs from the familiar parametrisation that underlies
ordinary spherical harmonics based upon the coset $SO(d-1)/SO(d-2)=S^{d-2}$.
We
 may
relate these two by focusing on the vector harmonics.  After elimination of
$v^m{}_{\tilde a}$ and $v^m{}_+$ by the constraints (2.15) and use of the
$SO(d-2)$ and $\{K^{\tilde m}\}$ gauge freedoms, the remaining light-like
$v^m=v^m{}_-$ may be written as $v^m=(R,Rv^{\hat m})$, $\hat m=1,\ldots,d-1$,
where $v^{\hat m}v^{\hat m}=1$.  The $v^{\hat m}$ can then be identified
with the usual vector spherical harmonics.  Higher spherical harmonics can
then be obtained as symmetrised traceless products of the $v^{\hat m}$.  The
problem of constructing general Lorentz covariant functions in terms of the
$v^m{}_a$ is the same as that of constructing conformally covariant
functions on $S^{d-2}$.  An essential feature of this construction is the
use of functions with a homogeneous scaling behaviour under the
$SO^\uparrow(1,1)$ symmetry.  The construction of general conformally
covariant functions in this way has been treated in ref.\ [16].
In the present paper, we shall not discuss further the construction of general
Lorentz-covariant functions  necessary for field-theoretical
applications, but shall restrict our attention to the invariance properties of
particle actions such as (2.1).

We end this section with the following  comment. One may  relax the
positivity conditions on $v^0$ (2.3) and $d\tau'/d\tau$, by requiring instead
that they be nonvanishing. This  corresponds to an equivalent representation
for
the  coset space (2.7)
$$
S^{d-2}={SO(1,d-1)\over[SO(d-2)\times SO(1,1)]\semidirprod
\{K^{\tilde m}\}}.\eqno(2.16)
$$
However, the above positivity conditions naturally appear in the superparticle
theory, to which we now turn.
\bigskip
\noindent{\bf 3. The superparticle in $d=2+1$}
\medskip
     For the superparticle, the coset (2.7) is still the space
that the $v^m$ take their values in, but in order to accommodate the
superparticle's spinor variable, we now make
use of the fundamental spinor representation of the Lorentz group. In
other words, we will consider the  smallest coset space of the $\hbox{\it
Spin}\,(1,d-1)$ group, which is the covering group of
$SO^\uparrow(1,d-1)$.  Since the spinor representations grow
exponentially in $d$ compared to the vector representation used above,
there will now be a dimensional dependence in our treatment.
 We
begin in $d=2+1$, where  $\hbox{\it Spin}\,(1,2)=SL(2,\R)$ and the coset (2.7)
is
represented starting from a real unimodular $2\times 2$ matrix
$e^\mu{}_\alpha \in SL(2,\R)$, where the greek indices denote spinor
representations and run over the values $1,2$. As in the bosonic case
above, the $SL(2,\R)$ transformations act by left multiplication,
$$
 e'^\mu{}_\alpha=\Lambda^\mu{}_\nu e^\nu{}_\alpha,\eqno(3.1)
$$
where $\Lambda\in SL(2,\R)$, so $\det\Lambda=1$.  The maximal subgroup ${\cal
H}_B$ of $SL(2,\R)$ (which in this case is known as the Borel subgroup) is
given
by the upper-triangular matrices $$
h=\pmatrix{a&b\cr
0&a^{-1}\cr},\ \ \ \ \ \ \ \ \ \ a\ne0\ \ \ a,b\in\R.\eqno(3.2)
$$
(Strictly speaking, this ${\cal H}_B$ differs from
$SO^\uparrow(1,1)\semidirprod \{K\}$ by a $Z_2$ factor, corresponding to
the possibility of both positive and negative values for $a$ in (3.2).  We
shall return to the issue of the double coverings below.)  The right-hand
index $\alpha$ corresponds to a reducible representation of ${\cal H}_B$,
with gauge transformations
$$
\eqalignno{ \psi'^{\mu}&=a \psi^{\mu}&(3.3a)\cr
\phi'^{\mu}&=a^{-1} \phi^{\mu}+b \psi^{\mu}&(3.3b)\cr}
$$
where $\psi^{\mu}=e^{\mu}{}_1$, $\phi^{\mu}=e^{\mu}{}_2$.

     The coset space (2.7) in this $d=2+1$ case is $S^1$.  Since this
one-dimensional space has fewer components than are contained in the
$e^\mu{}_\alpha$, the number of independent components of this matrix must be
reduced to just one.  This happens as a result of two particular features of
our construction.  The first is that the matrix $e^\mu{}_\alpha$ is
required to be an element of $SL(2,\R)$, and so there is a constraint.  In
terms of the ($\psi$, $\phi$) variables introduced above, this is, from
$\det(e^\mu{}_\alpha)=1$,
$$
\psi^1\phi^2-\psi^2\phi^1=1.\eqno(3.4)
$$
We also have the restriction that $\psi^\mu\ne0$, since this would
otherwise be inconsistent with $\det(e^\mu{}_\alpha)=1$.  This means that
either $\psi^1\ne0$ or $\psi^2\ne0$.  These cases define for us the two
charts needed to cover $S^1$.
Consider to be specific the first case.  In this case, we may solve the
constraint (2.4) for $\phi^2$: $$
\phi^2={1\over\psi^1}-\psi^2{\phi^1\over\psi^1}.\eqno(3.5)
$$
We now have three remaining degrees of freedom.  Two of these may be
removed using the second particular feature of our construction, namely the
${\cal H}_B$ gauge transformations
$$\eqalignno{
\phi'^1&=a^{-1}\phi^1+b\psi^1&(3.6)\cr
\psi'^\mu&=a\psi^\mu.&(3.7)\cr}
$$
Thus, we may pick, {\it e.g.}, the gauge $\phi^1=0$, and take $x=\psi^2/\psi^1$
as a gauge invariant coordinate in this chart. In the other chart, where
$\psi^2\ne0$, one proceeds analogously but dividing by $\psi^2$, and ending up
with the remaining coordinate $y=\psi^1/\psi^2$. In the overlapping region of
these charts $x\ne0,\;y\ne0$ and $xy=1$, so $x$ and $y$ can be viewed as
stereographic coordinates on the circle, arising from projection from the north
and south poles onto the equatorial line.

     In order to clarify the discussion of the higher-dimensional cases
to follow, we now consider in more detail the relation between the spinor
representation construction of (3.1--3.4) and the vector representation
construction based on $SO^\uparrow(1,2)$ (see (2.9),(2.13)). The spinor and
vector  representation matrices  are related by
 $$
e^\mu{}_\alpha
e^\nu{}_\beta\sigma^m_{\mu\nu}=v^m{}_a\sigma_{\alpha\beta}^a,\eqno(3.8)
$$
where
$$
\sigma^m_{\mu\nu}=\pmatrix{1&0\cr0&1\cr},\ \  \pmatrix{0&1\cr1&0\cr},\ \
\pmatrix{-1&0\cr0&1\cr},\ \ \ \ m=0,1,2. \eqno(3.9)
$$
Given $e^\mu{}_\alpha$, we find $v^m{}_a$ as
$$
v^m{}_a=-\ft12e^\mu{}_\alpha
e^\nu{}_\beta\sigma^m_{\mu\nu}\tilde\sigma^{\alpha\beta}_a, \eqno(3.10)
$$
where $\tilde\sigma^{m\mu\nu}
= \epsilon^{\mu\mu'}\epsilon^{\nu\nu'}\sigma^m_{\mu'\nu'}$,
and $\tr(\sigma^m\tilde\sigma^n)=-2\eta^{mn}$. Note that $v^0{}_0$ is strictly
positive, $v^0{}_0 = \ft12((e^1{}_1)^2 + (e^1{}_2)^2 + (e^2{}_1)^2 +
(e^2{}_2)^2)>0$.

 One can easily check that the unimodularity constraint on $e^\mu{}_\alpha$
is equivalent to the requirement that $v^m{}_a\in SO^\uparrow(1,2)$ and that
the spinorial matrix $e^\mu{}_\alpha$ is defined by the vectorial matrix
$v^m{}_a$ up to a sign, in accordance with
$$
{SL(2,\R)\over Z_2} = {{\it Spin}(1,2)\over Z_2} =SO^\uparrow(1,2). \eqno(3.11)
$$

     Although we have already shown that our coset space (2.7) in the $d=2+1$
case is the circle $S^1$, it is instructive to use the relation of the
spinor representation to
the vector representation (3.8) in order to see once again that the
manifold is $S^1$.   In the $d=2+1$ case, the transverse $\tilde a$ index
in (2.7) can take only the value $\tilde a=1$.  Thus, we need to
construct the $v^m{}_-$, $v^m{}_1$ and $v^m{}_+$ vectors from the
$\psi^\mu$ and $\phi^\mu$ spinors. Rewriting  the lower $a$ index  in the
light-cone basis (2.11)  we get $$\eqalignno{
v^m{}_-&= \ft1{\sqrt2}\sigma^m_{\mu\nu}\psi^\mu\psi^\nu&(3.12a)\cr
v^m{}_1&=\sigma^m_{\mu\nu}\psi^\mu\phi^\nu&(2.12b)\cr
v^m{}_+&=\ft1{\sqrt2}\sigma^m_{\mu\nu}\phi^\mu\phi^\nu.&(3.12c)\cr}
$$
It is important to emphasize that in such a representation
$v^m{}_-$ is automatically a null vector: it is expressed in terms of a
two-component spinor $\psi^\mu$, and it is impossible to construct a Lorentz
scalar from $\psi^\mu$ alone.  From the spinor representation gauge
transformations (3.3), one  recovers the vector representation gauge
transformations (2.13).  Then, by our previous discussion for the vector
representation, we know that the manifold $SL(2,R)/{\cal H}_B$ is a circle.
Since ${\cal H}_B/Z_2={\cal H}=SO^\uparrow(1,1)\semidirprod \{K\}$ and
$SL(2,\R)/Z_2=SO^\uparrow(1,2)$, this spinor realisation of the coset describes
the same manifold $S^1$ as the vector construction (2.7) (since the $Z_2$
factors ``cancel out'').

     The main benefit of the spinor formulation is that it
allows for a natural interpretation of the $\kappa$-symmetry of the
superparticle as a local worldline supersymmetry. In particular, this
permits the $\kappa$ symmetry to be reformulated with an algebra that closes
without the use of equations of motion.  Staying for the moment with the
$d=2+1$ case, and following [3], we consider the target space coordinates
$x^m(\tau)$ and $\theta^\mu(\tau)$ as the lowest components of superfields
 $X^m$,
$\Theta^\mu$ on an $N=1$  worldline superspace with coordinates $(\tau,\eta)$.
As usual, we may expand these superfields into component fields using a
superspace covariant derivative
$$
D=\partial_\eta + \im\eta\partial_\tau,\eqno(3.13)
$$
which satisfies
$$
D^2=\im\partial_\tau.\eqno(3.14)
$$
For an unrestricted local worldline supersymmetry, it
would be necessary to include a ``super einbein'' in the construction of
$D$, but this einbein can always be gauged away, leaving $D$ in the form
(3.13). The residual covariance of $D$ is then restricted to
transformations of the form
$$\eqalign{
\delta\tau&=\Lambda-\ft12\eta
D\Lambda=\lambda(\tau)+\im\eta\epsilon(\tau)\cr
\delta\eta&=-\ft{\im}2D\Lambda=
\epsilon(\tau)+\ft12\eta\dot\lambda(\tau),\cr}\eqno(3.15)
 $$
where the components of the superfield parameter $\Lambda(\tau,\eta)=
\lambda(\tau)+2\im\eta\epsilon(\tau)$  correspond to ordinary worldline
diffeomorphisms and  local worldline supersymmetry. By the
residual covariance of (3.13) under (3.15), we mean that $D$ transforms into a
factor times itself, {\it i.e.}\
$$
\delta D=-(D\delta\eta)D.\eqno(3.16)
$$
Eq.\ (3.16) defines a superconformal transformation of the
super worldline (see appendix).  Under the local  transformations
(3.15), $\Theta(\tau,\eta)$ transforms as a scalar.  The superfield
$\Theta(\tau,\eta)$ can be expanded into component fields as
$$
\theta^\mu(\tau)=\Theta^\mu\for_{\eta=0},\ \ \ \ \
\psi^\mu(\tau)=D\Theta^\mu\for_{\eta=0}.\eqno(3.17)
$$
As a consequence, the worldline supersymmetry transformations of
$\theta(\tau)$ are
$$
\delta\theta^\mu(\tau)=-\epsilon(\tau)\psi^\mu(\tau).\eqno(3.18)
$$
This transformation can be related to the familiar $\kappa$-symmetry if
we choose the parameter $\epsilon$ to be field-dependent,
$$
\epsilon=\kappa_\nu\psi^\nu\eqno(3.19)
$$
and recall, from (3.12a) with the identification
$v^{\mu\nu}=\tilde\sigma_m^{\mu\nu}v^m=v^{(\mu\nu)}{}_-$, that
$v^{\mu\nu}=\ft1{\sqrt2}\psi^\mu\psi^\nu$ so that the right hand
side of (3.18) contains a light-like vector that is proportional {\it on
shell} ({\it c.f.}\ eq.\ (2.3)) to $p^{\mu\nu}=\sigma_m^{\mu\nu}p^m$.
This reproduces the standard form of the $\kappa$-symmetry transformation
$$
\delta\theta^\mu(\tau)\sim p^{\mu\nu}\kappa_\nu.\eqno(3.20)
$$

Note that in making the field-dependent substitution (3.19), one
introduces a two-component spinor anticommuting parameter $\kappa_\mu$
which contains twice as many degrees of freedom as $\epsilon$. Thus it
is not surprising that  $\kappa_\mu$ is defined modulo its own
``pre-gauge'' transformations
$$
\delta\kappa_\mu=\xi(\tau)\psi_\mu.\eqno(3.21)
$$
If one were to insist on using only  objects with the standard
spin-statistics correlation (anticommuting half-integer spins and
commuting integer ones), it would be necessary to put $\xi(\tau)=\rho^\nu(\tau)
\psi_\nu$ so that $\delta\kappa_\mu\sim p_{\mu\nu}\rho^\nu$; this would start
an
infinite sequence of pregauge transformations.

     The superparticle action in the $d=2+1$ case has both the restricted local
worldline supersymmetry (2.15) and also a rigid target-space
superconformal symmetry [3]:
$$
I_{d=2+1}=-\im\int d\tau d\eta P_{\mu\nu}(DX^{\mu\nu}-\im D\Theta^{(\mu}
\Theta^{\nu)}).\eqno(3.22)
$$
The component expansions of $X^{\mu\nu}$ and $P^{\mu\nu}$ are obtained
as above by expanding using $D$,
$$\eqalign{
x^{\mu\nu}(\tau)=\sigma_m^{\mu\nu}x^m(\tau)&=X^{\mu\nu}\for_{\eta=0}\cr
p^{\mu\nu}(\tau)&=P^{\mu\nu}\for_{\eta=0},\cr}\eqno(3.23)
$$
while the higher components of both of these superfields are auxiliary.
Upon elimination of the auxiliary fields in (3.22), one obtains
$$
I=\int d\tau p_{\mu\nu}(\dot
x^{\mu\nu}-\psi^\mu\psi^\nu-\im\dot\theta^{(\mu}\theta^{\nu)}).\eqno(3.24)
$$
Since one has eliminated auxiliary fields to obtain this form of the
action, the worldline supersymmetry transformations form a closed
algebra at this stage only modulo the equations of motion.  In order to
recover the conventional superparticle action, one should vary (3.24) with
respect to $\psi^\mu$, obtaining
$$
p_{\mu\nu}\psi^\mu=0,\eqno(3.25)
$$
which may in turn be solved by
$$
p_{\mu\nu}=e^{-1}\psi_\mu\psi_\nu \eqno(3.26)
$$
since $\psi_\mu$ is nonvanishing. Substituting for $\psi^\mu$ then
reproduces the usual form of the action
$$
\int d\tau[p_{\mu\nu}(\dot x^{\mu\nu}-\im\dot\theta^{(\mu}\theta^{\nu)}) -
\ft12ep_{\mu\nu}p^{\mu\nu}].\eqno(3.27)
$$

     Subject to the equations of motion following from (3.22), there is a
rigid part of the $d=1$ worldline supersymmetry (3.15)
that may be viewed as a projection of the target spacetime
supersymmetry.  This rigid worldline supersymmetry may also be defined as the
transformations under which the superspace derivative $D$ (3.13) is invariant.
In the projection, $D$ may be viewed as a ``spinorial
pull-back'' of the target superspace derivative $D_\mu$.  This projection
will help us to clarify why the worldline supersymmetry has half the number of
generators of the spacetime supersymmetry, which in this case yields an
unextended $N=1$ worldline supersymmetry.

     The target superspace derivative is
$$
D_\mu={\partial\over\partial\Theta^\mu}+\im\Theta^\nu{\partial\over\partial
X^{\mu\nu}},\eqno(3.28) $$
and satisfies the algebra
$$\eqalign{
\{D_\mu,D_\nu\}&=2\im D_{\mu\nu}\cr
[D_{\mu\nu},D_\rho]&=0,\cr}\eqno(3.29)
$$
where $D_{\mu\nu}=\partial/(\partial X^{\mu\nu})$.  Now in order to make
an identification between the worldline supersymmetry and a projected
version of this algebra, we will want to use the chain rule together with a
{\it specific} solution of the superparticle equations of motion, taken in an
appropriate gauge for the local worldline transformations (3.15).  The
equations of motion following from (3.22) expand into component fields to give
$$\eqalignno{
\dot p_{\mu\nu}&=0&(3.30a)\cr
\dot
x^{\mu\nu}&=\psi^\mu\psi^\nu+\im\dot\theta^{(\mu}\theta^{\nu)}&(3.30b)\cr
p_{\mu\nu}\psi^\nu&=0&(3.30c)\cr
p_{\mu\nu}\dot\theta^\mu&=0.&(3.30d)\cr}
$$
The solution to (3.30c) was given in (3.26); at
this point it is convenient to fix the $\tau$ reparametrisation gauge
in (3.15) by setting
$$
\dot e=0;\eqno(3.31)
$$
(3.30a) then implies
$$
\psi^\mu=\psi^\mu_{(0)}={\rm constant}.\eqno(3.32)
$$
Eq.\ (2.30d) then implies
$$
\dot\theta^\mu=\psi^\mu_{(0)}\zeta(\tau),\eqno(3.33)
$$
where $\zeta(\tau)$ is an arbitrary Grassmann function, but one may then use
the worldline  supersymmetry freedom $\epsilon(\tau)$ in (3.18) to set
$$
\xi(\tau)=0,\eqno(3.34)
$$
so
$$
\theta^\mu(\tau)=\theta_{(0)}^\mu={\rm constant}.\eqno(3.35)
$$
Finally, we may solve the $x^{\mu\nu}$ equation (3.30b)
by
$$
x^{\mu\nu}(\tau)=x^{\mu\nu}_{(0)}+
\psi^\mu_{(0)}\psi^\nu_{(0)}\tau.\eqno(3.36)
$$

     The covariant derivatives on the target superspace can then be
projected onto the worldline using $\psi^\mu_{(0)}$:
$$
\eqalignno{
D&=\psi^\mu_{(0)} D_\mu&(3.37)\cr
\partial_\tau&=\psi^\mu_{(0)}\psi^\nu_{(0)} D_{\mu\nu},&(3.38)\cr}
$$
with the resulting algebra
$$\eqalignno{
D^2&=\im\partial_\tau&(3.39a)\cr
[D,\partial_\tau]&=0.&(3.39b)\cr}
$$
The differential operators $D$ and $\partial_\tau$ are understood to act on
functions of $(X(\tau,\eta),\Theta(\tau,\eta))$, where the worldline
component fields of $X(\tau,\eta)$ and $\Theta(\tau,\eta)$ are given
in (3.32, 3.35, 3.36).  The gauge choices (3.31) and (3.34) leave unfixed a
rigid $N=1$ worldline supersymmetry, under which the worldline
superspace derivative $D$ (3.37) is invariant.

     In addition to the projected supersymmetry transformations, in
the $d=2+1$ case there is a scale transformation automorphism symmetry of the
$D$, $\partial_\tau$ algebra (3.16) that can be identified with the
$SO^\uparrow(1,1)$ part of the Borel subgroup ${\cal H}_B$.  In fact, one
may say that the whole of ${\cal H}_B$ acts on the algebra (3.14), but
that the derivatives $D$ and $\partial_\tau$ are inert under the
$K_{\tilde m}$ transformations.  This is easy to see using the projection
relations (3.37, 3.38), since $\psi^\mu_{(0)}$ is invariant under
$K_{\tilde m}$ as is evident from (3.3).

     The appearance of $N=1$ worldline supersymmetry in the projected
algebra of the operators (3.37, 3.38) is illustrative of the general
situation, in which the worldline supersymmetry has half the number of
components of the target spacetime supersymmetry.  This is well-known from
light-cone considerations. The analysis of the present paper  amounts to a
Lorentz covariantisation of the light-cone; the halving of the number of
supersymmetries on the worldline carries over directly.  This can be seen
already from a group-theoretical standpoint, for of the two spinors
potentially available to make projections of $D_\mu$ only $\psi^\mu$
transforms homogeneously under ${\cal H}_B$ transformations: as one can
see again from (3.3), $\phi^\mu$ transforms into $\psi^\mu$ under the
$K_{\tilde
 m}$
transformation.  As a result, only $\psi^\mu$ may be used to make a
$K_{\tilde m}$-invariant projected worldline derivative (3.37) so that
there is only one type of ``$\partial_\tau$'' derivative occurring in the
closure of the $D$ algebra, that given in (3.38).  If we had tried to
start from $\phi^\mu D_\mu$ (or to include it into the algebra together
with $\psi^\mu D_\mu$), closure of the ``worldline'' $D$ algebra
together with the ${\cal H}_B$ transformations would then lead to three
independent bosonic derivatives, $\psi^\mu\psi^\nu D_{\mu\nu}$,
$\psi^\mu\phi^\nu D_{\mu\nu}$ and $\phi^\mu\phi^\nu D_{\mu\nu}$. In other
words, there would be no projection onto the worldline, since effectively
all three of the target spacetime derivatives would appear.
\bigskip
\noindent{\bf 4. $d=3+1$}
\medskip
     In the case of the superparticle in $d=3+1$ dimensions, as is well
known, the ${\it Spin}(3,1)$ covering group of the Lorentz group is
isomorphic to $SL(2,\C)$.  Accordingly, in order to obtain a spinorial
realisation of the coset (2.7) for this case, we begin with a unimodular
$2\times2$ complex matrix $e^\mu{}_\alpha\in SL(2,\C)$, thus satisfying
again the constraint
$$ \det(e^\mu{}_\alpha)=1.\eqno(4.1)
$$
 The  $SL(2,\C)$ Lorentz transformations act on this matrix from the
left, $$
 e'^\mu{}_\alpha=\Lambda^\mu{}_\nu e^\nu{}_\alpha,\eqno(4.2)
$$
where $\Lambda\in SL(2,\C)$. The maximal subgroup  ${\cal H}_B$ of $SL(2,\C)$
consists of the upper triangular complex matrices
$$
h=\pmatrix{a&b\cr
0&a^{-1}\cr},\ \ \ \ \ \ \ \ \ \ a\ne0\ \ \ a,b\in\C.\eqno(4.3)
$$
Under the group ${\cal H}_B$ acting by right multiplication, the
$\alpha$ index on $e^\mu{}_\alpha$ becomes reducible but not fully
reducible:  for
$$
e^\mu{}_\alpha=(\psi^\mu, \phi^\mu)\in SL(2,\C)\ \ \ \
\alpha=1,2\eqno(4.4)
$$
the ${\cal H}_B$ transformations are
$$
\eqalignno{ \psi'^{\mu}&=a \psi^{\mu}&(4.5a)\cr
\phi'^{\mu}&=a^{-1} \phi^{\mu}+b \psi^{\mu}.&(4.5b)\cr}
$$

     The coset space $SL(2,\C)/{\cal H}_B$ in this $d=3+1$ case is just the
sphere $S^2$.  One way to see this is by studying the charts
corresponding to the different solutions of (4.1) and their corresponding
gauge choices, as in (3.4-3.7). Another way is to relate, in analogy with
(3.8),
the above spinor representation to the vector one via the $\sigma$-matrix
invariance equation $$
e^\mu{}_\alpha e^{\dot\mu}{}_{\dot\alpha}\sigma^m_{\mu\dot\mu}=
v^m{}_a\sigma^a_{\alpha\dot\alpha},\eqno(4.6)
$$
where $e^{\dot\mu}{}_{\dot\alpha}={\overline {(e^\mu{}_\alpha)}}$ and
$\sigma^m_{\mu\dot\mu}$  are the usual Van der Waerden matrices.
{}From eq.\ (4.6) one finds the vector representation matrix
$$
v^m{}_a=-\ft12e^\mu{}_\alpha
e^{\dot\mu}{}_{\dot\alpha}\sigma^m_{\mu\dot\mu}
\tilde\sigma_a^{\alpha\dot\alpha}.\eqno(4.7)
$$
which belongs to $SO^\uparrow(1,3)$ provided $e^\mu{}_\alpha\in SL(2,\C)$.
Note that the sign of $v^0{}_0$ is automatically positive since $v^0{}_0$ is
given by a positive quadratic form in the components of $e^\mu{}_\alpha$ and
 $e^{\dot\mu}{}_{\dot\alpha}$. Furthermore, one can follow the pattern of
our $d=2+1$ discusion in section 3 by noting that the ${\cal H}_B$
transformations on $e^\mu{}_\alpha$ imply via (4.7) the ${\cal H}$
transformations on $ v^m{}_a$. Note that  the expression for the
light-like vector  $ v^m{}_-$ is now
$$
v^m{}_-=\ft1{\sqrt2}\sigma^m_{\mu\dot\mu}\psi^\mu\bar\psi^{\dot\mu}.
\eqno(4.8)
$$

 With the spinorial coset parameter $\psi^\mu$, we can project the
$d=3+1$ spacetime covariant derivative algebra
$$\eqalignno{
\{D_\mu,D_\nu\}&=0&(4.9a)\cr
\{D_\mu,\bar D_{\dot\nu}\}&=2\im\sigma^m_{\mu\dot\nu}\partial_m&(4.9b)\cr}
$$
onto the worldline, following the
pattern of our $d=2+1$ discussion (3.37, 3.38) (but with $\psi^\mu$ now
complex).  The pulled-back derivatives are:
$$
\eqalignno{
D&=\psi^\mu D_\mu&(4.10a)\cr
\bar D&=\bar\psi^{\dot\mu}\bar D_{\dot\mu}.&(4.10b)\cr
\partial_\tau&=
\psi^\mu\bar\psi^{\dot\nu}\sigma^m_{\mu\dot\nu}\partial_m.&(4.10c)\cr}
$$
The resulting algebra corresponds to a rigid $N=2$ worldline supersymmetry:
$$\eqalignno{
\{D,\bar D\}&=2\im\partial_\tau&(4.11a)\cr
D^2&=\bar D^2=0.&(4.11b)\cr}
$$

     The worldline algebra (4.11) possesses an automorphism algebra
that, like the $d=2+1$ case of section 3, contains scale
transformations that may be identified with the $SO^\uparrow(1,1)$ part
of ${\cal H}_B$.  In the $d=3+1$ case, however, there is also a $U(1)$
automorphism that acts on ($D$, $\bar D$); this may be identified with the
$SO(2)$ part of ${\cal H}_B$.  As in the $d=2+1$ case, the $K^{\tilde m}$
transformations act trivially on ($D$, $\bar D$, $\partial_\tau$), as one
can see from (4.5) and (4.10).

     The rigid $N=2$ worldline supersymmetry may be generalised directly to
the local $N=2,\ d=1$ superconformal one (see appendix).
   The simplest way to describe the latter is not in terms of a real superspace
$\R^{(1\vert2)}: (\tau, \eta, \bar\eta)$, but in terms of a chiral worldline
superspace [6] $\C^{(1\vert1)}: (\tau_{\ssm L}, \eta)$, where
$$
\tau_{\ssm L}=\tau+\im\eta\bar\eta.\eqno(4.12)
$$
The restricted worldline supersymmetry then consists of the transformations
$$\eqalign{
\delta\tau_{\ssm L}=\Lambda-\bar\eta\bar
D\Lambda=\lambda(\tau_{\ssm L})+2\im\eta\bar\epsilon(\tau_{\ssm L}),\cr
\delta\eta=-\ft{\im}2\bar
D\Lambda=\epsilon(\tau_{\ssm L})+\eta(\ft12\dot\lambda+\im
\rho(\tau_{\ssm L}).\cr}\eqno(4.13) $$ The parameters $\lambda,\epsilon$ and
$\rho$ correspond respectively to worldline reparametrisations, local
supersymmetry and local $U(1)$ transformations.

     As in section 3, we may now relate the superspace
transformations (4.13) to $\kappa$-symmetry transformations.  Here, the
worldline superfields containing the component variables $x^m$ and
$\theta^\mu$ are now chiral superfields $X_{\ssm L}^m(\tau_{\ssm L}, \eta)$ and
$\Theta^\mu(\tau_{\ssm L}, \eta)$.  Following the pattern of (3.17), one
 identifies
the $\eta=0$ component of $\Theta^\mu(\tau_{\ssm L}, \eta)$ with
$\theta^\mu(\tau)$; the next higher component is then identified with the
$d=3+1$ complex commuting spinor variable $\psi^\mu$ of (4.4).  The
worldline supersymmetry transformation is then
$\delta\theta^\mu(\tau)=-\epsilon(\tau)\psi^\mu(\tau)$ (like (3.18), except
that $\theta$, $\epsilon$ and $\psi^\mu$ are now all complex).  One may then
choose the parameter $\epsilon(\tau)$ to take the field-dependent value
$\epsilon(\tau)=\bar\kappa_{\dot\nu}(\tau)\bar\psi^{\dot\nu}(\tau)$ as in
(2.19).  With this parameter, the variation of $\theta^\mu(\tau)$ takes the
form $\delta\theta^\mu =
\psi^\mu\bar\psi^{\dot\nu}\bar\kappa_{\dot\nu}$.  On shell,
$\psi^\mu\bar\psi^{\dot\nu}$is proportional to
$p^{\mu\dot\nu}$, so that using the equations of motion these transformations
give rise to the standard form of a $\kappa$-transformation.

     The equation of motion needed to make the identification of
$\kappa$-symmetry with the local worldline supersymmetry,
$$
p_m\sigma^m_{\mu\dot\nu}\bar\psi^{\dot\nu}=0,\eqno(4.14)
$$
follows directly from the superfield form of the $d=3+1$
superparticle equations of motion for $X_{\ssm L}^m(\tau_{\ssm L}, \eta)$ and
$\Theta^\mu(\tau_{\ssm L}, \eta)$:
$$
\eqalignno{
\ft{\im}2(X_{\ssm L}^m-X_{\ssm R}^m)+\Theta\sigma^m\bar\Theta&=0&(4.15a)\cr
\bar DP_m&=0&(4.15b)\cr
\bar D(P_m\sigma^m_{\mu\dot\nu}\bar\Theta^{\dot\nu})&=0,&(4.15c)\cr}
$$
where $P_m=P_m(\tau, \eta, \bar\eta)$ is a real $N=2$ superfield and $X^m_{\ssm
 R}$
is the (antichiral) complex conjugate of $X^m_{\ssm L}$.  The component
equation
(4.14) then follows directly from (4.15b) and (4.15c). The superfield
equations (4.15) may be derived from the $d=3+1$ superparticle action [6]
$$
I_{d=3+1}=\int d\tau d\eta d\bar\eta
P_m[\ft\im2(X^m_{\ssm L}-X^m_{\ssm R})
+ \Theta^\mu\sigma^m_{\mu\dot\nu}\bar\Theta^{\dot\nu}].\eqno(4.16)
$$
This action is invariant both under the local $N=2,\ d=1$ worldline
superconformal transformations (4.14) and also under rigid $N=1$, $d=3+1$
target spacetime superconformal symmetry.
\bigskip
\noindent{\bf 5. $d=5+1$}
\medskip
     In $d=5+1$, the ${\it Spin}(1,5)$ covering group of the Lorentz group
is isomorphic to $SL(2,\H)$.  This group can be defined as the group of
complex unimodular $4\times4$ matrices $e^\mu{}_\alpha$ subject to the
pseudoreality condition
$$
\overline{e^\mu{}_\alpha}=C^{\dot\mu}{}_{\nu}e^\nu{}_\beta
C^{-1\beta}{}_{\dot\alpha}\eqno(5.1)
$$
where $C$ is the $d=5+1$ charge conjugation matrix.  We work here with the
chiral (fundamental) representation.  In $d=5+1$, there are no ordinary
Majorana spinors, but since complex conjugation does not produce an
inequivalent representation in this dimension, linear combinations of spinors
and their complex conjugates are covariant.  In this way, one may rewrite an
ordinary Weyl spinor as an $SU(2)$-Majorana spinor, satisfying
$$
\overline{\psi^\mu{}_i}=C^{\dot\mu}{}_\nu
\epsilon^{ij}\psi^\nu{}_j.\eqno(5.2)
$$
In this basis, the charge conjugation matrix takes the form
$$
C^{\dot\mu}{}_\nu=\pmatrix
{0&-1&0&0\cr
1&0&0&0\cr
0&0&0&-1\cr
0&0&1&0\cr}.\eqno(5.3)
$$

     In constructing the coset (1.6) in the $d=5+1$ case using the spinor
representation, the maximal subgroup ${\cal H}_B$ is
$[SU(2)\times\widetilde{SU(2)}\times SO(1,1)]\semidirprod\{K_{\tilde m}\}$.  As
in the $d=2+1$ and $d=3+1$ cases, the ${\cal H}_B$ subgroup may be represented
by block upper triangular matrices, where the blocks are themselves $2\times2$
matrices -- the quaternions of $SL(2,\H)$.  When decomposed with respect to the
${\cal H}_B$ subgroup, the index $\alpha$ on the matrix $e^\mu{}_\alpha$
becomes reducible:
$$
e^\mu{}_\alpha=(\psi^\mu{}_{\ssm A}, \phi^\mu{}_{\dot {\ssm A}}).\eqno(5.4)
$$
where $\ssm A$ and $\dot{\ssm A}$ correspond respectively to $SU(2)$ and
$\widetilde{SU(2)}$. The spinors $\psi^\mu{}_{\ssm A}$ and $\phi^\mu{}_{\dot
{\ssm A}}$ in (5.4)  satisfy the $SU(2)$-Majorana condition (5.2) and the whole
matrix $e^\mu{}_\alpha$ must be unimodular.  In terms of $\psi^\mu{}_{\ssm A}$
 and
$\phi^\mu{}_{\dot {\ssm A}}$, the unimodularity condition becomes
$$
\epsilon_{\mu\nu\rho\sigma}\psi^\mu{}_{\ssm A}\psi^\nu{}_{\ssm B}
\phi^\rho{}_{\dot {\ssm C}}\phi^\sigma{}_{\dot {\ssm
D}}=\epsilon_{{\ssm A}{\ssm B}}\epsilon_{\dot {\ssm C}\dot {\ssm
D}}.\eqno(5.5)
$$
It follows from the unimodularity condition (5.5) that neither
$\psi^\mu{}_{\ssm A}$ nor $\phi^\mu{}_{\dot{\ssm A}}$ can vanish.

     The ${\cal H}_B$ gauge transformations of the components of (5.4) are
given by
$$\eqalignno{
\delta\psi^\mu{}_{\ssm A}
&=\omega\psi^\mu{}_{\ssm A} + \omega_{({\ssm A}{\ssm
B})}\epsilon^{{\ssm B}{\ssm C}}\psi^\mu{}_{\ssm C}&(5.6a)\cr
\delta\phi^\mu{}_{\dot{\ssm A}} &=-\omega\phi^\mu{}_{\dot{\ssm A}} +
\omega_{\dot {\ssm A}\dot {\ssm B}}\epsilon^{\dot {\ssm B}\dot
{\ssm C}}\phi^\mu{}_{\dot {\ssm C}} + k_{{\ssm A}\dot
{\ssm A}}\epsilon^{{\ssm A}{\ssm B}}\psi^\mu{}_{\ssm B}.&(5.6b)\cr}
$$
Upon use of the unimodularity condition (5.5) and elimination of gauge
components using the $\omega_{\dot {\ssm A}\dot {\ssm B}}$ and $k_{{\ssm A}\dot
{\ssm A}}$
transformations in (5.6$b$), all of the components of
$\phi^\mu{}_{\dot\ssm A}$ may be considered to be dependent. The
independent components then live in $\psi^\mu{}_{\ssm A}
$; but of them half still
transform under the $SU(2)$ and $SO(1,1)$ scale transformations.  Instead of
eliminating three components of $\psi^\mu{}_{\ssm A}$ using the $SU(2)$, one
may
instead make the $SU(2)$ gauge-invariant construction
$$
v^{\mu\nu}=\psi^\mu{}_{\ssm A}\psi^\nu{}_{\ssm B}\epsilon^{{\ssm A}{\ssm
B}}.\eqno(5.7)
$$
This antisymmetric object in two spinorial indices is equivalent to an
$SO(1,5)$ vector,
$$
v^m=\gamma_{\mu\nu}^mv^{\mu\nu},\eqno(5.8)
$$
where $v^m$ is a real vector as a consequence of the $SU(2)$-Majorana condition
(5.2) on $\psi^\mu{}_{\ssm A}$.  In terms of $v^{\mu\nu}$, this
becomes a pseudoreality condition $\overline{v^{\mu\nu}}=C^{\dot\mu}{}_\rho
C^{\dot\nu}{}_\tau v^{\rho\tau}$. The vector $v^m$ is also lightlike, since
$$
v^mv_m=-2v^{\mu\nu}v^{\rho\sigma}\epsilon_{\mu\nu\rho\sigma}=0,\eqno(5.9)
$$
as can be seen from the fact that the ${\ssm A}$ index on the
commuting variable $\psi^\mu{}_{\ssm A}$ runs over two values only.  Moreover,
since $\psi^\mu{}_{\ssm A}$ is nonvanishing, the same holds for $v^m$.
Finally,
since we have not yet fixed the $SO(1,1)$ scale transformations, the vector
$v^m$ is only defined up to a scale.  At this point, we have established in
the $d=5+1$ case the conditions for the vector $v^m$ that were shown in
section 2 to define a sphere, so that in the $d=5+1$ case, $v^m$ and hence
$\psi^\mu{}_{\ssm A}$ parametrise the sphere $S^4$.

   The spinorial coset parameter $\psi^\mu{}_{\ssm A}$ can be used, following
 the
pattern of our earlier discussions, to project the $d=5+1$ spacetime
covariant derivative algebra onto the worldline.  The rigid spacetime
covariant derivative algebra here is
$$
\{D^i_\mu,D^j_\nu\}=2\im\epsilon^{ij
}\gamma^m_{\mu\nu}\partial_m,\eqno(5.10)
$$
where the indices $i,j$ correspond to the $SU(2)_{\scriptscriptstyle {\rm
aut}}$
outer automorphism of the rigid $d=5+1$ supersymmetry algebra.  In analogy
with the lower dimensional cases, we can now use linearly
transforming $\psi^\mu{}_{\ssm A}$ to make the projections
$$
\eqalignno{
D^i_{\ssm A}&=\psi^\mu_{\ssm A}D^i_\mu&(5.11a)\cr
\partial_\tau&=\epsilon^{{\ssm A}{\ssm B}}
\psi^\mu_{\ssm A}\psi^\nu_{\ssm B}\gamma^m_{\mu\nu}\partial_m.&(5.11b)\cr}
$$
 The resulting
algebra is the $N=4,\ d=1$ extended worldline covariant derivative algebra
$$
\{D^i_{\ssm A},D^j_{\ssm
B}\}=-\im\epsilon^{ij}\epsilon_{{\ssm A}{\ssm B}}\partial_\tau.\eqno(5.12) $$
This algebra has an automorphism group
$SU(2)\times SU(2)_{\scriptscriptstyle {\rm aut}}\times SO(1,1)$, where the
first $SU(2)$ operates on the ${\ssm A},{\ssm B}$ indices and
$SU(2)_{\scriptscriptstyle {\rm aut}}$
operates on the $i,j$ indices.  Note that the former is a subgroup of
the Borel subgroup ${\cal H}_B$ (the $\omega_{({\ssm AB})}$
transformations in (5.6)), while the latter is inherited from the rigid
$SU(2)$ automorphism of the $d=5+1$ spacetime covariant derivative
algebra (5.10).

     The rigid $N=4$ worldline supersymmetry corresponding to the algebra
(5.12) can then be promoted to a local worldline supersymmetry algebra, the
$N=4,\ d=1$ superconformal algebra, in accordance with our earlier
discussions.  The new feature of the $d=5+1$ case in distinction to our
earlier discussions is that the $SU(2)_{\scriptscriptstyle {\rm aut}}$ group
now plays a significant r\^ole.  The worldline superspace coordinates
corresponding to (5.12) are $(\tau,\eta^i_{\ssm A})$.  The
$N=4,\ d=1$ superconformal transformations may be defined by the property that
$D^i_{\ssm A}$ transforms homogeneously into a matrix contracted with
itself, but does not transform into $\partial_\tau$.  These
transformations are given by
$$\eqalignno{
\delta\tau=&\Lambda+\ft12\eta^i_A D_i^A\Lambda&(5.13a)\cr
\delta\eta^i_A=&-\im D^i_A\Lambda,&(5.13b)\cr}
$$
where $\Lambda(\tau,\eta)$ is an arbitrary real superfield parameter (see
appendix).

 The $(\tau,\eta^i_A)$ superspace $\R^{(1\vert4)}$ does not have any
invariant subspaces akin to $\C^{(1\vert1)}$ in the $d=3+1$ case, where
invariant here is understood to mean both under the $N=4,\ d=1$ supersymmetry
and also under the $SU(2)_{\scriptscriptstyle {\rm aut}}$, since the spinor
derivative $D^i_A$ carries an irreducible representation under the product of
these groups.

     Nonetheless, there is a way to find an invariant superspace with a
smaller fermionic coordinate than the $(\tau,\eta^i_{\ssm A})$ superspace
[6].  This involves, however, an initial extension of the superspace.  The
$(\tau,\eta^i_{\ssm A})$ superspace may be viewed as the group manifold of
the $N=4,\ d=1$ super Poincar\'e group.  Alternatively, since the $N=4,\
d=1$ super Poincar\'e group has an $SO(4)\simeq SU(2)\times
SU(2)_{\scriptscriptstyle {\rm aut}}$ automorphism, $\R^{(1\vert4)}$ may also
be viewed as the coset space ${\cal E}/{\cal F}$, where $\cal E$ is the $N=4,\
d=1$ super Poincar\'e group times $SU(2)\times SU(2)_{\scriptscriptstyle
{\rm aut}}$ and $\cal F$ is $SU(2)\times SU(2)_{\scriptscriptstyle {\rm
aut}}$.  The virtue of this reformulation is that it suggests another
coset space, which one may denote $\R_{\ssm H}^{1+2\vert4}$, defined as
the coset ${\cal E}/{\cal F'}$ where ${\cal F'}=SU(2)\times
U(1)_{\scriptscriptstyle {\rm aut}}$.  The notation $\R_{\ssm
H}^{1+2\vert4}$ means that this space has $1+2$ bosonic coordinates and
$4$ fermionic coordinates.  The two extra  bosonic coordinates come from
the extra $S^2=SU(2)_{\scriptscriptstyle {\rm
aut}}/U(1)_{\scriptscriptstyle {\rm aut}}$
 factor.

     Unlike $\R^{(1\vert4)}$, the extended superspace $\R_{\ssm H}^{1+2\vert4}$
does have an invariant subspace [6].  One can see this from the fact that
the supersymmetry generator $Q^i_{\ssm A}$ becomes reducible under the
smaller $\cal F'$ stability subgroup.  The way to define the invariant
subspace is to introduce harmonic coordinates $u^\pm_i$ for the
$S^2=SU(2)_{\scriptscriptstyle {\rm aut}}/U(1)_{\scriptscriptstyle {\rm aut}}$
factor space and then pass to the ``analytic basis'' variables
$$\eqalignno{
\hat\tau&=\tau+\ft{\im}2\eta^{{\ssm A}i}\eta_{\ssm A}^ju^+_i
u^-_j&(5.14a)\cr
\eta^{\ssm A\pm}&=\eta^{{\ssm A}i}u^{\pm}_i.&(5.14b)\cr}
$$
 One can show that the analytic subspace $\hat\tau,\;\eta^{{\ssm A}+},\;
u^\pm_i$ is invariant under the full superconformal group provided the
 harmonics transform as
$$
\delta u^+_i=\rho^{++}u^+_i, \;\; \delta u^-_i=0 \eqno(5.15)
$$
where $\rho^{++}=\ft{\im}2D^{{\ssm A}+}D^+_{\ssm A}\Lambda$.
In the analytic basis, the projected derivative $u^+_iD_{\ssm A}^i$
becomes a partial derivative $-\partial/\partial\eta^{{\ssm A}-}$.
On the other hand, the covariant harmonic derivative
$D^{++}=u^+_i\partial/\partial u^-_i$ takes the form
$$
D^{++}=u^+_i\partial/\partial u^-_i+ \ft{\im}2
\eta^{{\ssm A}+}\eta^+_{\ssm A}\partial_{\hat\tau}+
\eta^{{\ssm A}+}\partial/\partial\eta^{{\ssm A}-}.\eqno(5.16)
$$
In terms of the  analytic worldline superfields
$P_m(\hat\tau,\eta^+,u),\;X^m(\hat\tau,\eta^+,u),\;
\Theta^{\mu+}(\hat\tau,\eta^+,u)$ and $N^{++}(\hat\tau,\eta^+,u)$, one can
write down the superparticle equation of motion as
$$
D^{++}P_m=0, \; D^{++}X^m=i\Theta^+\gamma^m\Theta^+,\;
D^{++}N^{++}+(N^{++})^2=0. \eqno(5.17)
$$
These can be derived from an following action formulated in the analytic
superspace [6]
$$\eqalign{
S&=-i\int du\,d\hat\tau \,d\eta^{{\ssm A}+}d\eta^+_{\ssm A}\big[
P_m(D^{++}X^m-i\Theta^+\gamma^m\Theta^+)\cr
&\phantom{=-i\int du\,d\hat\tau \,d\eta^{{\ssm A}+}d\eta^+_{\ssm A}\big[
P_m(}+
P^-_\mu\Theta^{\mu +}+ P^{--}(D^{++}N^{++}+(N^{++})^2)
\big],\cr}\eqno(5.18)
$$
where $P^-_\mu$ and $P^{--}$ are analytic Lagrange multipliers. The
superfield $N^{++}$ plays the r\^ole of a bridge between the worldline
harmonics $u^{\pm}_i$ and the target space ones $w^{\pm}_i$: $$
w^+_i=u^+_i+N^{++}u^-_i,\;\;w^-_i=u^-_i. \eqno(5.19)
$$
The  gauge transformations of $N^{++}$ are given by $\delta
N^{++}=-\rho^{++}$, so the target space harmonics $w^\pm_i$ are inert under the
worldline superconformal group (5.15). The
two sets of harmonics do not coincide off shell. However, on shell the
superfield $N^{++}$ can be completely gauged away, thus identifying
$w^{\pm}_i$ with $u^{\pm}_i$. As a result of this gauge fixing,
$\rho^{++}=0$ and the superfield parameter $\Lambda$ of the $N=4$
superconformal group becomes constrained
$$
D^A_iD_{Aj}\Lambda=0. \eqno(5.20)
$$
Note that this restriction is consistent with the group composition
law ($A.8$).
\bigskip
\noindent{\bf 6. $d=9+1$}
\def\a{\alpha}
\def\b{\beta}
\def\c{\gamma}\def\C{\Gamma}
\def\d{\delta}

\def\h{\eta}

\def\t{\theta}\def\T{\Theta}

\def\m{\mu}
\def\f{\phi}
\def\n{\nu}

\def\ta{\tau}
\def\x{\chi}
\def\o{\omega}\def\O{\Omega}

\bigskip
We turn finally to the most interesting case, $d=10$. For $N=1$ spacetime
supersymmetry the spinors are Majorana-Weyl, {\it i.e.}\ sixteen-component
real Weyl spinors. The spin group, $Spin (1,9)$, is related to the Lorentz
group, $SO^\uparrow (1,9)$, via the invariance of the gamma-matrices; if
$e^\m{}_\a\in Spin(1,9)$ (so that it is a real $16\times 16$ matrix), one
has the constraint
$$
e^\m{}_\a e^\n{}_\b (\c^m)_{\m\n} = v^m{}_a (\c^a)_{\a\b}
\eqno(6.1)
$$
where $v$ is the corresponding element of $SO^\uparrow (1,9)$. The
$\gamma$-matrices in (6.1) are symmetric on their spinor indices; they are not
Dirac matrices, but are analogs of the $\sigma$-matrices in four dimensions.
Spinorial indices cannot be raised or lowered in ten dimensions, but there is
another set of matrices $\gamma_a^{\alpha\beta}$ with upper indices.\footnote
{$^{*}$}{\tenfoot These obey the anticommutation relation
 $\gamma^a_{\alpha\beta}\gamma^{b\beta\gamma}+
\gamma^b_{\alpha\beta}\gamma^{a\beta\gamma}=2\eta^{ab}\delta^\gamma_\alpha.$
In a specific choice of spinor basis one may have
$\gamma^a_{\alpha\beta}=\gamma_a^{\alpha\beta}$ numerically and also the
values
$$
\gamma^0_{\alpha\beta}=\left(\matrix{\oneone&0\cr 0&\oneone
\cr}\right),\;\; \gamma^{\tilde
a}_{\alpha\beta}=\left(\matrix{0&\sigma^{\tilde a}\cr
 (\sigma^{\tilde a})^{\ssm T}&0\cr}\right),\;\;
\gamma^9_{\alpha\beta}=\left(\matrix{-\oneone&0\cr 0&\oneone \cr}
\right).
$$
An important identity is $\gamma_{a\alpha(\beta}\gamma^a_{\gamma\delta)}=0$.}
Equation (6.1) can be  expanded in a complete set of symmetric $16\times 16$
matrices consisting of $(\c^a)_{\a\b}$ and $(\c^{abcde})_{\a\b}$, the latter
being self-dual on its vectorial indices. We get $$
e^\m{}_\a e^\n{}_\b (\c^m)_{\m\n} (\c_a)^{\a\b} = 16 v^m{}_a
\eqno(6.2)
$$
and
$$
e^\m{}_\a e^\n{}_\b (\c^m)_{\m\n} (\c^{abcde})^{\a\b} = 0.
\eqno(6.3)
$$
The matrix $e^\mu{}_\alpha$, considered as the product of ${\bf\overline{16}}$
and ${\bf 16}$ representations of $SO(1,9)$, is decomposable into
${\bf 1}+{\bf 45}+{\bf 210}$. Now, eq.\ (6.3) imposes precisely $210$
 conditions,
while eq.\ (6.2) is equivalent to the unimodularity condition $(\det e)=1$. The
remaining $45$ degrees of freedom correspond to parameters of $SO(1,9)$.

 The Borel subgroup $H_B=[SO(8)\times
SO^\uparrow(1,1)]\semidirprod \{K_{\tilde m}\}$ acts on the $\a,\b$ indices by
$$ e'^\m{}_\a= e^\m{}_\b \Omega^\b{}_\a.
\eqno(6.4)
$$
To describe this action explicitly, it is convenient to split the $\a$ index
up into its $SO(8)$ irreducible parts, $e^\m{}_\a=
(e^\m{}_{\ssm A},\  e^\m{}_{\dot\ssm A} ),$ where  ${\ssm A},\;{\dot{\ssm A}}=
1\ldots 8$  are
correspondingly indices of the  ${\bf 8_s}$ and ${\bf 8_c}$ fundamental spinor
representations of $SO(8)$. The matrix $\Omega$ is block triangular,
 $$
\Omega^\a{}_\b=\left(\matrix{\Omega^{\ssm A}{}_{\ssm B}&
\Omega^{\ssm A}{}_{\dot{\ssm B}}
\cr 0&\Omega^{\dot{\ssm A}}{}_ {\dot{\ssm B}}\cr}\right). \eqno(6.5)
$$
Here $\Omega^{\ssm A}{}_{\ssm B}$ is an $SO(8)$ matrix in the ${\bf 8_s}$
representation times an $SO(1,1)$ scale factor, $\Omega^{\dot{\ssm
A}}{}_{\dot{\ssm B}}$ is the same $SO(8)$ matrix in the ${\bf 8_c}$
representation times the opposite $SO(1,1)$ factor, {\it i.e.}\ infinitesimally
we have $$
\Omega^{\ssm A}{}_{\ssm B}\simeq \delta^{\ssm A}{}_{\ssm B}(1+\omega)+
\omega^{\ssm A}{}_{\ssm B},\;\;
\Omega^{\dot{\ssm A}}{}_{\dot{\ssm B}}\simeq \delta^{\dot{\ssm A}}{}_{\dot
{\ssm B}}(1-\omega)+ \omega^{\dot{\ssm A}}{}_{\dot{\ssm B}},\eqno(6.6)
$$
and the $K_{\tilde m}$ parameter $\Omega^{\ssm A}{}_{\dot{\ssm B}}$
corresponds  to the eight-dimensional vector ${\bf 8_v}$ representation of
$SO(8)$.

 Thus, under an $H_B$-transformation $e^\m{}_{\ssm A}$
transforms homogeneously  while
$e^\m{}_{\dot{\ssm A}}$ transforms into $e^\m{}_{\ssm A}$ as well as into
 itself. The
constraints (6.2), (6.3) together with the gauge transformations (6.4) imply
that the components of $e^\m{}_{\dot{\ssm A}}$ are either gauge degrees
of freedom or are
functions of the components of $e^\m{}_{\ssm A}$. We can thus focus on
$e^\m{}_{\ssm A}$. From (6.1) and the values given for $(\c^a)_{{\ssm
A}{\ssm B}}$ we have
$$
\eqalign{
e^\m{}_{\ssm A} e^\n{}_{\ssm B} (\c^m)_{\m\n} &= v^m{}_a (\c^a)_{{\ssm A}{\ssm
B}}=v^m{}_- (\c^-)_{{\ssm A}{\ssm B}}\cr &=\sqrt 2 v^m{}_- \d_{{\ssm A}{\ssm
B}}
 \cr}
\eqno(6.7)
$$
where we have introduced a light-cone decomposition of the vector index,
$\c^a=(\c^-,\c^{\tilde a},\c^+)$. The vector $v^m{}_-=\ft1{8\sqrt2}
e^\m{}_{\ssm A}
e^\n{}_{\ssm B} (\c^m)_{\m\n} \d^{{\ssm A}{\ssm B}}$ is lightlike and invariant
under the  $SO(8)\semidirprod\{K_{\tilde m}\} $ subgroup of $H_B$. The
independent coset space parameters are contained in the vector $v^m{}_-$
which is defined up to a positive scale transformation. Hence the coset
is the eight-sphere $S^8$.

The matrix $e^\m{}_{\ssm A}$ can be used to project from $d=10$ superspace onto
the worldline. The flat $d=10$ covariant derivatives $D_\m,\partial_m$
obey $$
\{D_\m,D_\n\} = 2i(\c^m)_{\m\n} \partial_m
\eqno(6.8)
$$
and projected covariant derivatives on the worldline can be defined by
$$
\eqalign{
D_{\ssm A}&= e^\m{}_{\ssm A} D_\m\cr \partial_\ta &=\ft18e^\m{}_{\ssm A}
e^\n{}_{\ssm A}  (\c^m)_{\m\n} \partial_m.\cr}
\eqno(6.9)
$$
The projected derivatives satisfy the $N=8,\; d=1$ extended worldline
covariant derivative algebra
$$
\{D_{\ssm A},D_{\ssm B}\} = 2i\d_{{\ssm A}{\ssm B}}\partial_\tau.
\eqno(6.10)
$$

It is  not known what the $N=8$ worldline supersymmetric action is for the
 $d=10$
superparticle, but solutions to the equations of motion can be written down in
a
way that is manifestly $N=8$ worldline supersymmetric. On the space ${\cal
M}\times Spin(1,9)$ (${\cal M}$=superspace), introduce new coordinates $x^a=
x^m
v^a{}_m,
 \t^\alpha
= \t^\m e^\alpha{}_\m$ together with the coset coordinates $e^\m{}_\alpha$,
 where
$v^a{}_m$
and $e^\alpha{}_\m$ are the inverses of $v^m{}_a$ and $e^\m{}_\alpha$. Then a
super-lightlike line in ${\cal M}$ passing through the origin is
described by the $H_B$-covariant equations
$$
x^+=x^{\tilde a} = \t^{\dot{\ssm A}} = 0.
\eqno(6.11)
$$
Equivalently, we can write
$$
\eqalign{
x^m &= x^- v^m{}_-\cr \t^\m&=\t^{\ssm A} e^\m{}_{\ssm A}\cr }
\eqno(6.12)
$$
with
$$
v^m{}_- = \ft1{8\sqrt2}e^\m{}_{\ssm A} e^\n{}_{\ssm A} (\gamma^m)_{\mu\nu}.
\eqno(6.13)
$$
Identifying the worldline parameters $(\ta,\h^{\ssm A})$ with $(x^+,
\t^{\ssm A})$ we
can see that (6.12) is a solution of the differential equations
$$
\eqalign{
D_{\ssm A} X^m - iD_{\ssm A} \T^\m (\c^m)_{\m\n} \T^\n &= 0\cr
D_{\ssm A} \T^\m &= e^\m{}_{\ssm A},\cr}
\eqno(6.14)
$$
where $X^m(\ta,\h^{\ssm A})$ and $\T^\m(\ta,\h^{\ssm A})$ are now $N=8$
worldline superfields and $e^\m{}_{\ssm A}$ is constant. Eq.\ (6.12)
gives the solution of (6.14) up to a (supersymmetric) shift of origin.
Note that these $d=10$ equations are not target space superconformally
invariant, although the corresponding equations of motion (and actions)
are for $d=3,4$ and $6$. If one defines the superconformal group of the
superspace as the group of transformations which scale the line element,
then in $d=10$ this group consists only of super-Poincar\'e and scale
transformations. In order to include conformal boosts it would be
necessary to extend the bosonic part of the algebra; such extensions
exist abstractly but cannot be realised on ordinary superspace.

It is interesting to relate our new formalism to  other approaches in the
literature. From (6.7) we can see that any of the $e^\m{}_{\ssm A},\
 {\ssm A}=1,\dots 8$
determines the same lightlike vector. Indeed, we can take the independent
parameters of the coset $S^8$ to be contained in the sixteen-component
spinor $e^\m{}_1 \equiv \psi^\m$. The space of $\psi^\m$ modulo scale
transformations is $S^{15}$ and is related to $S^8$ by the Hopf
fibration, which has fibre $S^7$. From the transformation (see
(6.4)--(6.6)) $$ \d e^\m{}_{\ssm A} = e^\m{}_{\ssm A}\omega
+e^\m{}_{\ssm B}\o^{\ssm B}{}_{\ssm A}
\eqno(6.15)
$$
 one finds
$$
\d \psi^\m =\psi^\m \omega+e^\mu{}_{\ssm A}\omega^{\ssm A}{}_{ 1}.
\eqno(6.16)
$$
The $SO(8)$ algebra matrix $\omega^{\ssm A}{}_{\ssm B}=
\omega_{{\ssm A}{\ssm B}}$ is antisymmetric, $\omega_{{\ssm A}{\ssm B}}$=
$-\ \omega_{{\ssm B}{\ssm A}}$. Hence
$\omega^{\ssm A}{}_{1}$ corresponds, in fact, to seven parameters
$\omega^{{\hat
{\ssm A}}}{}_{1},\; {{\hat{\ssm A}}}=2,\ldots 8$. Without losing  generality
one can make a field-dependent change of transformation parameter,
analogously to eq.\ (3.19), $$
\omega_{\hat{\ssm A}1}=
\xi_\nu e^\nu{}_{{\hat{\ssm A}}}
\eqno(6.17)
$$
where $\xi_\nu$ is an arbitrary  commuting $SO(1,9)$ spinor parameter.
Substituting this back into (6.15) and using the identities
$ e^\nu{}_{{\hat{\ssm A}}} e^\mu{}_{{\hat{\ssm A}}}=
 e^\nu{}_{\ssm A} e^\mu{}_{\ssm A}- e^\nu{}_{1} e^\mu{}_{1}$ and
$$
e^\nu{}_{\ssm A} e^\mu{}_{\ssm A}=\ft1{16}(\gamma_m)^{\mu\nu}
e^\rho{}_{\ssm A}(\gamma^m)_{\rho\sigma} e^\sigma{}_{\ssm A} \eqno(6.18)
$$
(which is just  equivalent to the condition $e^\m{}_\a e^\n{}_\b
(\gamma_a)^{\alpha\beta} = v^m{}_a (\gamma_m)^{\mu\nu}$ conjugate to
(6.1), taken in the $-$ projection of $a$) and also $$
e^\rho{}_{\ssm A}(\gamma^m)_{\rho\sigma} e^\sigma{}_{\ssm A}=8
e^\rho{}_{1}(\gamma^m)_{\rho\sigma} e^\sigma{}_{1}=
8\psi^\rho(\gamma^m)_{\rho\sigma}\psi^\sigma,\eqno(6.19)
$$
we finally get
$$
\delta\psi^\mu=\omega'\psi^\mu +\ft12(\gamma_m)^{\mu\nu}
\xi_\nu\psi\gamma^m\psi. \eqno(6.20)
$$
with the scale factor $\omega'=\omega-\xi_\nu\psi^\nu$.

The transformation
(6.20) is precisely that used by Berkovits in his approach to the superparticle
[7]. Note that as a result of the field-dependent change (6.17), the gauge
transformations (6.4) are realised in (6.20) nonlinearly. Moreover, the
 $16$-dimensional parameter $\xi_\nu$ introduced in (6.17) instead of the
$7$-dimensional one $\omega_{\hat{\ssm A}1}$ is defined also up to a
pregauge freedom [7,5]
$$
\delta\xi_\mu=(\gamma^m)_{\mu\nu}\psi^\nu r^m\eqno(6.21)
$$
with a $10$-dimensional parameter $r^m$ which is defined up to the
transformation
$$
\delta r^m=(\gamma^m)_{\mu\nu}\psi^\mu\psi^\nu q \eqno(6.22)
$$
where $q$ is a scalar parameter.

In fact, it is possible to generalise Berkovits' formalism in an $SO(8)$
covariant fashion by using $e^\mu{}_{\ssm A}$ instead of $\psi^\mu=e^\mu{}_1$.
One could then solve for $p^m$ as $p^m=e^\mu{}_{\ssm A}(\gamma^m)_{\mu\nu}
e^\nu{}_{\ssm A}$ and introduce the additional variables
$$\eqalign{
f_{\mu{\ssm A}}&=x^m(\gamma_m)_{\mu\nu}e^\nu{}_{\ssm A}+
\im (\gamma^m)_{\mu\nu}\theta^\nu e^\rho{}_{\ssm
A}(\gamma_m)_{\rho\sigma}\theta^\sigma,\cr
\lambda^m_{\ssm A}&=
\theta^\mu (\gamma^m)_{\mu\nu}e^\nu{}_{\ssm A}.\cr}\eqno(6.23)
$$
The Brink-Schwarz Lagrangian can then be rewritten in terms of the
constrained variables $(e^\mu{}_{\ssm A},\;f_{\mu{\ssm
A}},\;\lambda^m_{\ssm A})$.
Alternatively, one can use the inverse group element
$e^\alpha{}_\mu=(e^{\ssm A}{}_\mu,\;e^{\dot\ssm A}{}_\mu)$ and write
$p^m=e^{\dot\ssm A}{}_\mu(\gamma^m)^{\mu\nu} e^{\dot\ssm A}{}_\nu$. The
additional variables are then $$\eqalign{
f^{\mu{\dot\ssm A}}&=x^m(\gamma_m)^{\mu\nu}e^{\dot\ssm A}{}_\nu +
\im \theta^\mu\theta^\nu e^{\dot\ssm A}{}_\nu,\cr
\lambda^{\dot\ssm A}&=\theta^\mu e^{\dot\ssm A}{}_\mu,\cr}\eqno(6.24)
$$
the complete final set of constrained variables being $(e^{\dot\ssm
A}{}_\mu,\;f^{\mu{\dot\ssm A}}, \;\lambda^{\dot\ssm A})$.  This option
does not exist in $d=3,4$ where a Berkovits-type formalism would differ
from the supertwistor one solely in its choice of a Grassmann-odd
variable instead of a Grassmann-even one. In $d=6$, one choice of
chirality in the solution for $p^m$ leads to a Berkovits-type formalism,
whereas the other choice gives rise to a supertwistor formalism [17].

Finally, we note that our Lorentz-coset formalism also allows for
pure spinors. If we set $\psi^\m= e^\m{}_1$ and $\f^\m=e^\m{}_2$ it is easy to
show that $\x^\m = \psi^\m + i\f^\m$ is pure, {\it i.e.}\ it obeys the
 constraint
$$
\x^\m (\c^m)_{\m\n} \x^\n = 0
\eqno(6.25)
$$
It has been suggested that pure spinors might play a r{\^o}le in
ten-dimensional supergeometry as it is possible to understand the constraints
of supergravity and super Yang-Mills theories as arising from integrability
along ``pure spinor lines''[12]. The space of pure spinors is the twenty-two
dimensional coset $SO^{\uparrow}(1,9)/SU(4)\semidirprod \{K_{\tilde m}\}$
[18]. This space is non-compact, but the space of projective pure spinors,
$SO^\uparrow(1,9)/ [U(4)\times SO^\uparrow(1,1)]\semidirprod\{K_{\tilde
m}\}$, is a compact space with twenty real dimensions. Locally this space
can be viewed as $S^8\times SO(8)/ U(4)$; globally it is the bundle of
complex structures preserving the metric on $S^8$ for which $SO(1,9)$
acts as the conformal group. (These complex structures are only locally
defined since $S^8$ is not a complex manifold.) This picture leads to the
possibility that the worldline coset for $N=8$ might be $SO(8)/U(4)$, and
that the target space coset might be extended to the space of projective
pure spinors. This would be a  natural extension of the $N=4$ case, since
there $SU(2)/U(1) = SO(4)/U(2)$.
\np
\leftline{\bf 7. Conclusions}
\medskip
\def\R{\rlap I\mkern3mu{{\rm R}}}
\def\H{\rlap I\mkern3mu{{\rm H}}}
\def\C{\mkern1mu\raise2.2pt\hbox{$\scriptscriptstyle|$}\mkern-7mu{{\rm C}}}
\def\O{\mkern1mu\raise2.2pt\hbox{$\scriptscriptstyle|$}\mkern-7mu{{\rm O}}}
In this paper we have interpreted the additional
``twistor-like'' variables occurring in the STV version of the
superparticle as parameters of the coset spaces $SO^\uparrow(1,d-1)/{\cal
H}_B$. We have seen that this description is valid for $d=3,4,6$ at the
level of the action and also for $d=10$ on-shell. Although we have
focused on flat target spaces, it is straightforward to extend this
formalism to include non-trivial supergravity and super-Maxwell
background fields [6]. In that case,  demanding that the projected
derivatives satisfy a flat superalgebra on the worldline puts constraints
on the backgrounds, and these constraints correspond to those derivable
from lightlike integrability [10,11]. For example, in the $d=10$ Maxwell
case, lightlike integrability implies the equations of motion for the
Maxwell multiplet.

The STV variables appear in the $\eta$-expansion of the worldline
spinorial superfields $\Theta$, and have not been introduced into the
target space as independent variables.  Although it should be possible
to extend the target space to incorporate them independently, it has not
been necessary to do so in the $d=3,4\&6$ cases in order to have a
satisfactory off-shell formulation, as we have seen. Indeed, although
we introduced both worldline and target space harmonic
variables in the $d=6$ case (following [6]), it is not obligatory to do so
since the model can be formulated off-shell in terms of constrained 4+4
component worldline $N=4$ superfields. On the other hand, it is likely
that additional variables will be necessary for the $d=10$ case. Since
there is no internal symmetry group in $d=10$, these new variables should
parametrise some coset of the Lorentz group. In this paper we have seen
that a candidate coset is the eight-sphere, which has the advantage of
being compact, in contradistinction to the previously proposed light-cone
harmonic superspace [14,15]. However, as we have remarked, there are
also other compact coset spaces such as the space of projective pure
spinors that could play a r\^ole in the off-shell formulation of the
$d=10$ superparticle.

Although we have not succeeded in giving a fully satisfactory
off-shell account of the $d=10$ superparticle, we believe that the compact
coset space approach will be of central importance to this problem. The
basic idea behind the STV and supertwistor approaches to the
superparticle is to solve the constraint $p^2=0$ by using spinor
variables. If this is done with a single minimal spinor $\psi,\  (p\sim
\psi^2)$, then the space of lightlike vectors up to a scale can be
interpreted in terms of the Hopf fibrations, {\it i.e.}\ the spheres of
real spinors up to a positive scale $S^1, S^3, S^7, S^{15}$ are fibre
bundles over the celestial spheres $S^1, S^2, S^4, S^8$ with fibres
$1,S^1,S^3,S^7$. The main difference in the $d=10$ case is that $S^7$ is
not a group. This is an advantage of the coset approach, which guarantees
that the divisor gauge transformations form a group. Actually, as we have
seen, the $Spin(1,d-1)/{\cal H}_B$ coset can be parametrised by the first
column of a group element in $d=3,4$ and by a pseudo-real spinor for
$d=6$, so that in these three cases we get back directly to the Hopf
fibration picture. However, in $d=10$, we need the $16\times 8$ matrix
$e^{\mu}{}_{\ssm A}$ which is constrained by (6.7), (6.18) so that the
situation is somewhat different.

Another way of looking at the Hopf fibrations is via the division algebras.
The celestial spheres can be presented as $KP^1$ with $K=\R,\C,\H,\O$ for
$d=3,4,6,10$ respectively. The division algebra approach to supersymmetry in
these dimensions has been studied by many authors (see for instance [19,17])
and
in some sense the supergeometries for both the worldline and the target
superspaces can be considered as real, complex and quaternionic for $d=3,4,6$
respectively. The open problem of $d=10$ can therefore be thought of as the
problem of finding out how octonionic supergeometries should be defined. This
problem is certainly relevant to finding the $d=10$ action, for in $d=4$ and
$d=6$ it is necessary
to compexify and ``quaternionify'' (or rather ``harmonise'') the even
part of the target superspaces in order to go off shell.

Finally, we note that although the superparticle has proved to be a
surprisingly difficult and interesting problem, its solution should be
seen as only a step on the way to understanding the supergeometry of
ten-dimensional superspace and of the Green-Schwarz
superstring\footnote{$^*$}{\tenfoot In a recent interesting paper [20],
the $\kappa$-symmetry of the $d=4$ Green-Schwarz superstring is explained
as a world-sheet supersymmetry.}.
We believe that the coset method presented here may be useful tool for tackling
these questions.  \bigskip \centerline{\bf Acknowledgements}
\bigskip
A.G. would like to thank sincerely F. Delduc, B. de Wit, O.V. Ogievetsky, V.I.
Ogievetsky and E. Sokatchev  for valuable discussions and the Theoretical
Physics group of Imperial College for its warm hospitality.

\bigskip
\centerline{\bf Appendix. Superconformal groups in one dimension}
\medskip
Here we give some details of the worldline superconformal groups. We
start with the $N$-extended Poincar\'e supersymmetry algebra in one
dimension
$$
\{Q_a,Q_b\}=2\delta_{ab}P, \;\;\;(a,b=1,2,\ldots,N) \eqno(A.1)
$$
It possesses an $O(N)$ automorphism group with $Q_a$ being a vector, and
$P$ a singlet. It can be realised in a real superspace $R^{(1|N)}$:
$(\tau,\eta^a)$ with a single bosonic coordinate $\tau$ and $N$ Grassmann
coordinates $\eta^a$. The covariant Grassmann derivatives
$$
D_a={\partial\over \partial\eta^a}+\im\eta_a\partial_\tau \eqno(A.2)
$$
satisfy the algebra
$$
\{D_a,D_b\}=2\im\partial_\tau. \eqno(A.3)
$$

Under the super worldline diffeomorphisms $\tau'=
\tau+\delta\tau(\tau, \eta)$,$\;\eta'=\eta+\delta\eta(\tau\eta)$, the
Grassmann derivatives $D_a$ will in general transform into
themselves and $\partial_\tau$:
$$
\delta D_a\equiv D'_a-D_a= -(D_a\delta\eta_b)D_b
+[\im (D_a\delta\eta_b)\eta_b+\im
\delta\eta_a-D_a\delta\tau]\partial_\tau. \eqno(A.4)
$$
Hence $D_a$ will transform homogeneously provided the following
constraint is imposed on the parameters $\delta\tau$ and $\delta\eta$
$$
\im (D_a\delta\eta_b)\eta_b+\im \delta\eta_a-D_a\delta\tau =0.\eqno(A.5)
$$
The corresponding constrained transformations are the superconformal
ones. The constraint ($A.5$) can be equivalently derived from the
requirement that the line element $ds=d\tau+\im\eta_a d\eta_a$ transform
into itself, {\it i.e.}\ $\delta ds\sim ds$.

The general solution to the constraint ($A.5$) is expressed in terms of a
real scalar superfield $\Lambda(\tau,\eta)$
$$
\delta\eta_a =-\ft{\im}2D_a\Lambda,\;\;\delta\tau=\Lambda
-\ft12\eta_aD_a\Lambda.
\eqno(A.6)
$$
Then the transformation laws of the line element and covariant
derivatives are given by
$$\eqalign{
\delta ds&=(\partial_\tau\Lambda)\; ds, \cr
\delta D_a&=-\ft12(\partial_\tau\Lambda) \;D_a
 +\ft{\im}4[D_a, D_b]\Lambda\; D_b \cr}.\eqno(A.7)
 $$

Furthermore, the commutator of two transformations ($A.6$) yields a
transformation withe the bracket parameter
$$
\Lambda_{\rm br}=\Lambda_2\partial_\tau\Lambda_1 -
\Lambda_1\partial_\tau\Lambda_2 +\ft{\im}2D_a\Lambda_1D_a\Lambda_2.
\eqno(A.8)
$$
Now, let us characterise some specific features of the cases
$N=1,2,4,8$.

For {\bf N=1} the  superfield $\Lambda(\tau,
\eta)=\lambda(\tau)+2\im\eta\epsilon(\tau)$ contains the irreducible (1+1)
representation of N=1 supersymmetry, with $\lambda$ and $\epsilon$ being
translation and supertranslation parameters.

For {\bf N=2} $\Lambda(\tau,\eta)$ is also irreducible. It describes
the (2+2) multiplet
$$
\Lambda=\lambda+2\im\eta_a\epsilon_a+\im \eta_1\eta_2\rho,
 \eqno(A.9)
$$
where $\rho(\tau)$ corresponds to local $SO(2)$ rotations. This group
possesses a natural holomorphic structure. To see this one should
pass to the ``eigenvectors'' of $SO(2)=U(1)$,
$$\eqalign{
\eta=\ft1{\sqrt 2}(\eta_1+\im\eta_2),\;&\;\bar\eta=\ft1{\sqrt
2}(\eta_1-\im\eta_2),\cr
D=\ft1{\sqrt2}(D_1-\im D_2)
= {\partial\over\partial\eta}+\im\bar\eta\partial_\tau,\;&\; {\bar
D}=\ft1{\sqrt 2}(D_1+\im
D_2)={\partial\over\partial\bar\eta}+\im\eta\partial_\tau.\cr}
\eqno(A.10)
$$
The derivative $\bar D$ is nilpotent, $\bar D^2=0$, so
there exists a basis in $R^{(1|2)}$ where it is simply a partial
derivative. This is the chiral basis
$$
\{\tau_{\ssm L}=\tau+\im\eta\bar\eta,\;\eta,\;\bar\eta\};\;\;\; (\bar D)_{\ssm
 L}=
{\partial\over\partial\eta}.\eqno(A.11)
$$
One  can check that the
chiral subspace $(\tau_{\ssm L},\;\eta)$ is closed under the superconformal
transformations ($A.6$); for instance, $\delta\eta= -\ft{\im}2\bar D
\Lambda,\;\;\bar D\;\delta\eta=0$.

In the {\bf N=4} case the scalar superfield
$$\eqalign{
\Lambda(\tau,\eta)=\lambda+2\im\eta_a\epsilon_a
+&\ft{\im}2\eta_a\eta_b\rho_{ab} +\eta_a^3\xi_a+\eta^4k,\cr
\eta_a^3=\ft1{3!}\epsilon_{abcd}\eta_b\eta_c\eta_d,\;&\;
\eta^4=\ft1{4!}\epsilon_{abcd}\eta_a\eta_b\eta_c\eta_d,\cr}\eqno(A.12) $$
describes the reducible (8+8) representation of $N=4$ supersymmetry. Its
irreducible part can be obtained by imposing a self-duality constraint
$$
[D_a,D_b]\Lambda=\ft12\epsilon_{abcd}[D_c,D_d]\Lambda.
\eqno(A.13)
$$
A nontrivial property of this constraint  is its consistency with the
group composition law ($A.8$), {\it i.e.}, the bracket parameter ($A.10$)
$\Lambda_{\rm br}$ satisfies the same constraint.
The constraint ($A.13$) has a transparent meaning: in the $N=4$ case the
automorphism group $SO(4)$ is (locally) equivalent to the product of two
$SU(2)$
groups, $SO(4)\sim SU(2)\times SU(2)_{\cal A}$, and the
transformations of one of them, say
$SU(2)_{\cal A}$, are fixed by ($A.13$).

Passing to the $SU(2)\times SU(2)_{\cal A}$ spinor
notation\footnote{$^\dagger$}{\tenfoot Here  ${\sm A}=1,2$ and $i=1,2$ are
indices of $SU(2)$ and $SU(2)_{\cal A}$ respectively. The spinor
indices are raised and lowered as usual with the help of the
$\epsilon_{\ssm AB}$ and  $\epsilon_{ij}$ symbols, e.g.
$\eta_{{\ssm A}i}=\epsilon_{\ssm AB}\epsilon_{ij}\eta^{{\ssm B}j}$.},
$\;\eta_a\rightarrow \eta_{{\ssm A}i}, D_a\rightarrow
D_{{\ssm A}i}={\partial/\partial\eta^{{\ssm
A}i}}+\ft12\im\eta_{{\ssm A}i}\partial_\tau$, one can rewrite the
commutation relations ($A.3$) as
$$ [D_{{\ssm A}i},D_{{\ssm B}j}]=2\im
\epsilon_{\ssm AB}\epsilon_{ij}\partial_\tau \eqno(A.14)
$$ and the
constraint ($A.13$) as
$$
D_i^{\ssm A}D_{{\ssm A}j}\Lambda=0\eqno(A.15)
$$

Owing to the presence of the two $SU(2)$ automorphism groups, the algebra
($A.14$) is very similar to the algebra of $N=2, d=6$ supersymmetry. As
in the latter case, it is useful to introduce $SU(2)_{\cal A}/U(1)$
harmonics $u^\pm_i,\; u^{+i}u^-_i=1$ in order to convert all explicit
$SU(2)_{\cal A}$ indices into $U(1)$ charges: $D_{Ai}\rightarrow
D^\pm_A=u^+_iD^i_A, \; \eta^{Ai}\rightarrow \eta^{A\pm}= u^+_i\eta^{Ai}$.
In the extended superspace $\tau,\eta, u$ one can then pass to an analytic
basis
$$
\hat\tau =\tau+\ft{\im}2\eta^{Ai}\eta_A^ju^+_iu^-_j,\;\eta^{A\pm}=
u^+_i\eta^{Ai},\;u^\pm_i\eqno(A.16)
$$
and find that the extended superspace contains an analytic subspace  of
smaller Grassmann dimension with coordinates
$$
z_A=(\hat\tau,\;\eta^{A+},\; u)\eqno(A.17)
$$
The analytic superfields $\Phi(\hat\tau,\eta^{A+}, u)$ automatically satisfy
the constraint $D^+_{\ssm A}\Phi=0$, because in the analytic basis
($A.16$) $D^+_{\ssm A}$ becomes just $-\partial/\partial\eta^{{\ssm
A}-}$.

 An important property of the analytic subspace ($A.17$) is its closure
under the full unconstrained $N=4$ superconformal group ($A.6$), provided
that the harmonics transform as follows
$$
\delta u^+_i= \rho^{++}u^-_i,\; \delta
u^-_i=0,\;\;\rho^{++}=\ft{\im}2D^{A+}D^+_A\Lambda.\eqno(A.18) $$
Then the other variations are analytic as well,
$$
\delta\hat\tau= -\ft14D^{A+}D^+_A(\eta^{B-}\eta^-_B\Lambda),\;
\delta\eta^+_A=\ft{\im}2D^{B+}D^+_B(\eta^-_A\Lambda), \eqno(A.19)
$$
because the third power of $D^+$ vanishes automatically.

The analytic parameters are not arbitrary, but are constrained to satisfy
$$
D^{++}\rho^{++}=0,\;D^{++}\delta\hat\tau=\im\eta^{A+}\delta\eta^+_A,\;
D^{++}\delta\eta^+_A=\rho^{++}\eta^+_A. \eqno(A.20)
$$
Here $D^{++}=u^+_i\partial/\partial u^-_i
+\ft{\im}2\eta^{A+}\eta^+_A\partial_{\hat\tau}$ is one of the
covariant derivatives on the sphere $S^2=SU(2)_{\cal A}/U(1)$.  Note,
that the subgroup ($A.13$) is obtained simply by putting $\rho^{++}=0$.

An interesting feature of the transformations ($A.18$), ($A.19$) is that
they preserve the supervolume of the analytic subspace ($A.17$):
$$
{\partial\over\partial\hat\tau}\delta\hat\tau +u^-_i{\partial\over\partial
u^+_i}-{\partial\over\partial\eta^{A+}}\delta\eta^{A+}=0 \eqno(A.21)
$$

Finally we turn to the {\bf $N=8$} case. Here, one encounters difficulty
even defining an appropriate worldline superconformal algebra.
In this case, the general unconstrained superfield
$\Lambda(\tau,\eta)$ describing (128+128) degrees of freedom is a highly
reducible representation of $N=8$  supersymmetry. Following the $N=4$
case, it is natural to ask whether there are any representations aside
from the full $\Lambda$ itself, that are consistent with the group
composition law ($A.8$).

The only $SO(8)$ covariant
constraint one can impose on $\Lambda$ without putting it on shell is the
fourth-order self-duality constraint
 $$
D_aD_bD_bD_b\Lambda=\ft1{4!}\epsilon_{abcdefgh}D_eD_fD_gD_h\Lambda.
\eqno(A.22)
$$
However, this constraint is not compatible with ($A.8$). Therefore in
order to constrain $\Lambda$ one would have to explicitly break the
tangent group $SO(8)$ symmetry by restricting the
$SO(8)$ parameter $[D_a,D_b]\Lambda$. Now one can
easily prove a lemma: if the surviving tangent group is a subgroup of
$SO(7)$, then this would put $\Lambda$ on shell, {\it i.e.},
such a restriction would be too strong. Indeed, the constraint
corresponding to a surviving $SO(7)$ tangent group reads
 $D_1D_r\Lambda=0$, where $r=2,\ldots,8$ (the 28 parameters of $SO(8)$ are
decomposed with respect to $SO(7)$ as {\bf 21}+{\bf 7} and the {\bf 7}
is set to zero). However, acting on this constraint with $D_1$ one
would get the on-shell condition $\partial_\tau D_r\Lambda=0$ and hence
an unwanted restriction on the ordinary worldline supersymmetry.
Therefore the sought-for tangent group should not be a subgroup of
$SO(7)$. This, in particular, excludes all of the simple proper subgroups
of $SO(8)$. We will not go into further detail here, but mention only
that the breaking of $SO(8)$ down to a nonsimple subgroup $SO(6)\times
SO(2)$ that has been suggested in the literature also puts $\Lambda$ on
shell.  Thus, whether it is possible to formulate an appropriate $N=8$
worldline superconformal algebra for the superparticle remains an open
problem.
\bigskip\bigskip

\centerline{\bf REFERENCES}
\bigskip

\item{[1]} W. Siegel, {\sl Phys.\ Lett.}{\bf 128B}(1983)397;
{\sl Class.\ Quantum Grav.}\ {\bf 2} (1985) 302.
\item{[2]} E. Bergshoeff, R. Kallosh and A. Van Proeyen, {\it
Superparticle actions and gauge fixings,} CERN preprint TH.6020/91 \ (to be
published in {\sl Class.\ Quantum Grav.}).
\item{[3]} D.P. Sorokin, V.I. Tkach and D.V. Volkov, {\sl Mod.\ Phys.\
Lett.}\ {\bf A4} (1989) 901;\nl
D.P. Sorokin, V.I. Tkach, D.V. Volkov and
A.A. Zheltukhin, {\sl Phys.\ Lett.}\ {\bf 216B} (1989) 302.
\item{[4]} R. Brooks, F. Muhammad and S.J. Gates, {\sl Class.\ Quantum
Grav.}\ {\bf 3} (1986) 745;\nl
J.Kowalski-Glikman,  J.W. van Holten, S. Aoyama and
J. Lukierski, {\sl Phys.\ Lett.}\ {\bf 201B} (1988) 487.
\item{[5]} P.S. Howe and P.K. Townsend, {\sl Phys.\ Lett.}\ {\bf 259B}
(1991) 285.
\item{[6]} F. Delduc and E. Sokatchev, {\it Superparticle with
extended worldline superpsymmetry}, preprint PAR-LPTHE/91-14 \ (to
be published in {\sl Nucl.\ Phys.}\ {\bf B}).
\item{[7]} N. Berkovits, {\sl Phys.\ Lett.}\ {\bf 247B}(1990) 45.
\item{[8]} R. Penrose and W. Rindler, {\it Spinors and space-time}
(Cambridge University Press, 1986).
\item{[9]}  A. Galperin, E. Ivanov, S. Kalitzin, V. Ogievetsky and
E. Sokatchev, {\sl Class.\ Quantum Grav.}\ {\bf 1} (1984) 469.
\item{[10]} E. Witten, {\sl Nucl.\ Phys.}\ {\bf B266} (1986) 245.
\item{[11]} E. Bergshoeff, P. Howe, C.N.\ Pope, E. Sezgin and
E. Sokatchev, {\sl Nucl.\ Phys.}\ {\bf B354} (1991) 113.
\item{[12]} P.S. Howe, {\sl Phys.\ Lett.}\ {\bf 258B} (1991) 141.
\item{[13]}  E. Bergshoeff, F. Delduc and E .Sokatchev, {\sl Phys.\
Lett.}\ {\bf 262B} (1991) 444;\nl
P.S. Howe, {\it Pure spinors, function superspaces and
supergravity theories in ten and eleven dimensions}, Stony Brook preprint
ITP-SB-91-18.
\item{[14]}  E. Sokatchev, {\sl Phys.\ Lett.}\ {\bf 169B} (1986) 209;
{\sl Class. Quantum Grav.} {\bf 4} (1987) 237.
\item{[15]}  E. Nissimov, S. Pacheva and S. Solomon, {\sl Nucl.\ Phys.}\
{\bf B296} (1988) 462;\nl
E. Nissimov, S. Pacheva and S. Solomon, {\sl Nucl.\ Phys.}\
{\bf B297} (1988) 349;\nl
R. Kallosh and M. Rahmanov, {\sl Phys.\ Lett.}\ {\bf 209B} (1988) 233.
\item{[16]}  I. Gel'fand, M. Graev and N. Vilenkin, {\it Generalized
functions}, v.5, (Academic Press, New York and London, 1966);\nl
N. Vilenkin, {\it
Special functions and the theory of group representations} (American
Mathematical Society, Providence, 1968).
\item{[17]}I. Bengtsson and M. Cederwall,
{\sl Nucl.\ Phys.}\ {\bf B221} (1988) 81.
\item{[18]}  P. Furlan and R. Raczka, {\sl J.\ Math.\ Phys.}\ {\bf
26} (1985) 3021.
\item{[19]} T. Kugo and P. Townsend, {\sl Nucl.\ Phys.}\ {\bf B221}
(1983) 357.
\item{[19]} E.A. Ivanov and A.A. Kapustnikov, {\it Towards a tensor
calculus for $\kappa$-supersymmetry}, preprint IC/91/68.
\bye